 \documentclass[final,3p,12pt,times]{elsarticle}

\usepackage{amssymb}
\usepackage{subcaption}
\usepackage{amsmath}
\usepackage{graphicx}
\usepackage{color}

\journal{arXiv}

\begin{document}

\begin{frontmatter}



\title{A method for signal components identification in acoustic signal with non-Gaussian background noise using clustering of data in time-frequency domain. }


\author[label1]{Anita Drewnicka}
\author[label1]{Anna Michalak} 
\author[label1]{Radosław Zimroz}
\author[label2]{Anil Kumar\corref{cor1}} \author[label3]{Agnieszka Wyłomańska} \author[label1]{Jacek Wodecki}
\affiliation[label1]{organization={Faculty of Geoengineering, Mining and Geology, Wrocław University of Science and Technology},
            addressline={Na~Grobli~15}, 
            city={Wrocław},
            postcode={50-421}, 
            country={Poland}}
 \affiliation[label2]{organization={College of Mechanical and Electrical Engineering, Wenzhou University},
            city={Wenzhou},
            postcode={325 035}, 
            country={China}}
\affiliation[label3]{organization={Faculty of Pure and Applied Mathematics, Wrocław University of Science and Technology},
            addressline={Wybrzeże Wyspiańskiego~27}, 
            city={Wrocław},
            postcode={50-370}, 
            country={Poland}}

\begin{abstract}
The problem of damage detection based on noisy vibration/acoustic signal analysis is well understood by researchers when the observed noise comes from the Gaussian distribution. However, if there are random impulsive and wideband disturbances in the measured data, the task becomes more difficult. High-amplitude non-cyclic impulsive disturbances can occur in real-life signals due to some random aspects of the industrial process, i.e. non-uniform character of the process over time, random impacts, etc. In this paper, the authors propose to analyze the distribution densities of the vectors of individual spectra that come from the spectrogram. A simple additive model consists of the signal of interest (SOI), Gaussian, and non-Gaussian noise. The random non-cyclic component is given by the three-point distribution. A clustering approach using the density-based spatial clustering algorithm (DBSCAN) was proposed to isolate separate classes of vectors of individual spectra from the spectrogram that are supposed to represent the occurrence of events that belong to different components of the analyzed signal. Such an approach allows the separation of different behaviors in the signal and the isolation of the information related to the fault. To verify the effectiveness of the proposed method, the envelope spectrum-based indicator (ENVSI) was used. The method was successfully tested on a real signal measured on an industrial machine with a faulty bearing.
\end{abstract}




\end{frontmatter}




\date{} 
\section{Introduction}
\label{introduction}

Vibration-based diagnostics is a common approach for detecting local faults in rotating machines \cite{zoubir2017contribution, wodecki2020separation, michalak2019application}. A time-frequency representation, such as a spectrogram, is often used because it allows the analysis of signals with a complex frequency structure over time. In some cases, cyclic impulsive variability in the spectral content of the signal can indicate the presence of damage, which is a common approach to local damage detection. Such methods are very helpful in predictive maintenance, where early detection and correction can prevent catastrophic failure or downtime \cite{yan2014wavelets, anwarsha2022review, he2022bearing, liang2022intelligent, wodecki2020separation}. 

The frequency range associated with the damage signature is often identified by selecting an informative frequency band (IFB). Various statistics are commonly used as filter characteristics, known as IFB selectors.
One of the most popular selector is spectral kurtosis, which is based on the calculation of the kurtosis value for each frequency bin in the spectrogram to obtain the filter \cite{antoni2007cyclic,antoni2006spectral}.
Many extensions, inspired by spectral kurtosis, have been invented, such as selectors based on the parameters of probability distributions \cite{hebda2020informative}, other statistics \cite{obuchowski2014selection}, dependency measures \cite{nowicki2021dependency}. 

Throughout the years, numerous variations and expansions have emerged of the IFB-based local damage detection method. These methods rely on a discrete frequency spectrum division. The kurtogram proposed by Antoni~\cite{antoni2007fast} stands out as the most popular among them. It is based on the concept of analyzing the kurtosis of a signal over different frequency bands. As with selectors, kurtogram extensions involve using other statistics instead of kurtosis. 
Antonis proposed an infogram that relies on measuring the negentropy of both the squared envelope and the squared envelope spectrum \cite{antoni2016info}. Zhao et al. introduced the Harsogram~\cite{zhao2016detection}, which uses the Harmonic Significant Index (HSI). A sparsity index, also known as the Gini index, is proposed as an alternative approach to resonance band selection in the work of Miao et al.~\cite{miao2017improvement_GINI}. Bozchalooi et al.~\cite{bozchalooi2007smoothness} present a method that focuses on the spectral smoothness index. This is defined as the ratio of the geometric mean to the arithmetic mean of the moduli of the Gabor wavelet coefficients. The autogram, introduced by Moshrefzadeh et al.~\cite{Moshrefzadeh2018294}, is constructed using the autocorrelation of the squared envelope spectrum. As an alternative method for spectrum division, Wang et al.~\cite{Wang2022} proposed the TIEgram, which is based on EES analysis. 
The diagnostic feature indicator was used to assess the richness of information regarding bearing faults within a specified frequency range in the IFBI$_\alpha$-gram \cite{schmidt2020methodology}. DTMSgram employs methods to reduce noise in both the time and frequency domains, aiming to improve the detection of informative frequency bands~\cite{Liu2024}.

Classical methods assume the Gaussian nature of the signal background. This assumption appears in many scientific works \cite{obuchowski2014selection,pancaldi2023impact,antoni2006spectral,saufi2017review}. Unfortunately, local damage detection in real-life scenarios is a much more complex task, as exemplified by the case of crushers or sieving screens, where falling rocks introduce a non-cyclic high-amplitude impulsive component in addition to the cyclic impulses generated by the fault \cite{michalak2020model, hebda2020informative}. It means that the noise occurring in the observed data is non-Gaussian and has an impulsive character, which violates the assumptions perpetuated by the classical methods and very often renders them ineffective. This problem has been noticed more and more frequently in recent years.

Smith et al. proposed a band selection method based on optimized spectral kurtosis for impulsive background noise with electromagnetic interference\cite{smith2016optimised}. Kruczek et al. explained how to analyze cyclostationarity in signals with complicated spectral structure \cite{kruczek2020detect}, and Wodecki et al. estimated how much impulsive contamination can be acceptable to detect fault using cyclostationary methods \cite{wodecki2021influence}. Wylomanska et al. introduced a denoising technique suitable for non-Gaussian signals \cite{wylomanska2016impulsive}. Borghesani et al. examined the impact of alpha-stable noise on traditional indicators \cite{borghesani2017cs2}, while Barszcz et al. addressed this problem using protrugram \cite{barszcz2011novel}. 

In general, basic selectors, such as commonly used spectral kurtosis, are ineffective in the presence of high-amplitude impulsive disturbances. However, some of the methods mentioned in the literature are still effective if the cyclic and random impulses do not overlap spectrally or if the spectral overlap is limited. In other words, there is a co-occurrence of the frequency characteristics of different impulse types. In addition, many approaches do not consider the spectral overlap of the informative and non-informative impulses \cite{hebda2020informative, miao2017improvement_GINI, antoni2016info}. When such overlap exists, more advanced methods are needed to identify and separate the components \cite{WODECKI2021108400}.

In recent years, neural network-based methods have gained significant popularity in bearing diagnostics, marking a shift toward data-driven approaches \cite{zheng2024progressive,ye2023intelligent, xu2022fault,wissbrock2024more}. Although these approaches offer many advantages, they also present challenges, such as the need for large datasets and significant computational resources.

The authors propose a procedure aimed at addressing the challenges related to non-Gaussian noise components in the signal. The goal of this approach is to separate the elements present in the signal, including cyclic impulses, non-cyclic high-amplitude disturbances, and noise. This method involves clustering spectral vectors in the time domain based on their empirical distribution patterns. For this purpose, the Density-Based Spatial Clustering of Applications with Noise (DBSCAN) algorithm was used \cite{ester1996density}. This methodology requires a two-stage classification, with the first stage focusing on the separation of high-amplitude, broadband, non-cyclic impulses. In order to extract the signal of interest from the noise, the remaining part of the data was reclassified and processed.

The rest of the paper is organized as follows: the methodology used in the further analysis was explained in Section \ref{sec2}. Section \ref{sec3} presents the results of the simulation and analysis. Section \ref{subsec3.2} presents the results of the test of the effectiveness of the method using real acoustic data measured in the laboratory and under industrial conditions. The effectiveness of the presented method based on Monte Carlo simulations is discussed in Section \ref{sec4}. Finally, the conclusions are formulated in Section \ref{sec5}.


\section{Methodology}\label{sec2}

In this Section the proposed procedure is presented.

\begin{figure}[ht!]
\includegraphics[width=0.35\textwidth, angle=0]{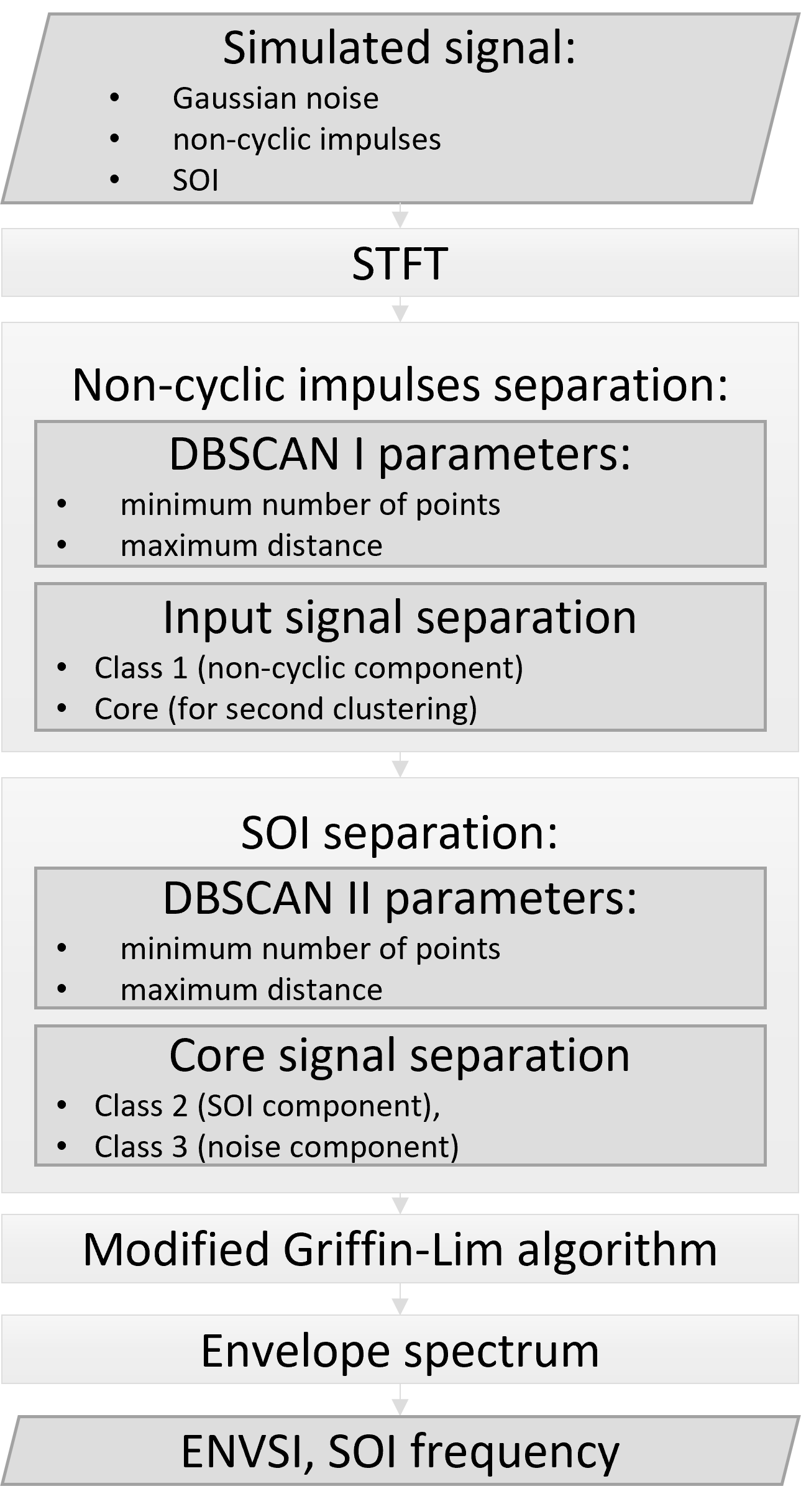}
\centering
\caption{Flowchart of the proposed procedure.}\label{schema1}
\end{figure}   
\unskip
In Figure \ref{schema1}, the flowchart of the proposed procedure is presented. Firstly, a signal consisting of Gaussian, non-Gaussian, and cyclic components corresponding to damage (SOI) was simulated. The signal was analyzed using a short-time Fourier transform (STFT) as a fundamental data representation. The subsequent column vectors of the spectrogram were treated as a spectral representation of the time window.

The proposed approach demonstrates a two-stage clustering process using DBSCAN, ensuring its adaptability to various signal analysis scenarios. 
In the first step, the Euclidean distance from the other vectors was calculated for each vector. Then, the signal was separated by using DBSCAN into two classes: non-cyclic impulses (referred to as class 1) and the core class, which was then forwarded for further analysis. In the second stage, the core class is clustered to form two classes: signal of interest (SOI) and noise.

Based on the classes of vectors analyzed obtained in this way, the signal was divided in the time domain into three classes corresponding to the subsequent components of the input signal (non-cyclic impulses, SOI, and noise). Since the proposed procedure uses double clustering, it has been named Double Spectral Clustering (DSC). 

Finally, time series are obtained for each class separately using a modified Gryffin-Lim algorithm. This allows for the calculation of the SES and the use of an indicator to evaluate the results.

\subsection{Signal construction}\label{s:sig}

\begin{figure}[ht!]
\centering
\includegraphics[width=0.75\textwidth, angle=0]{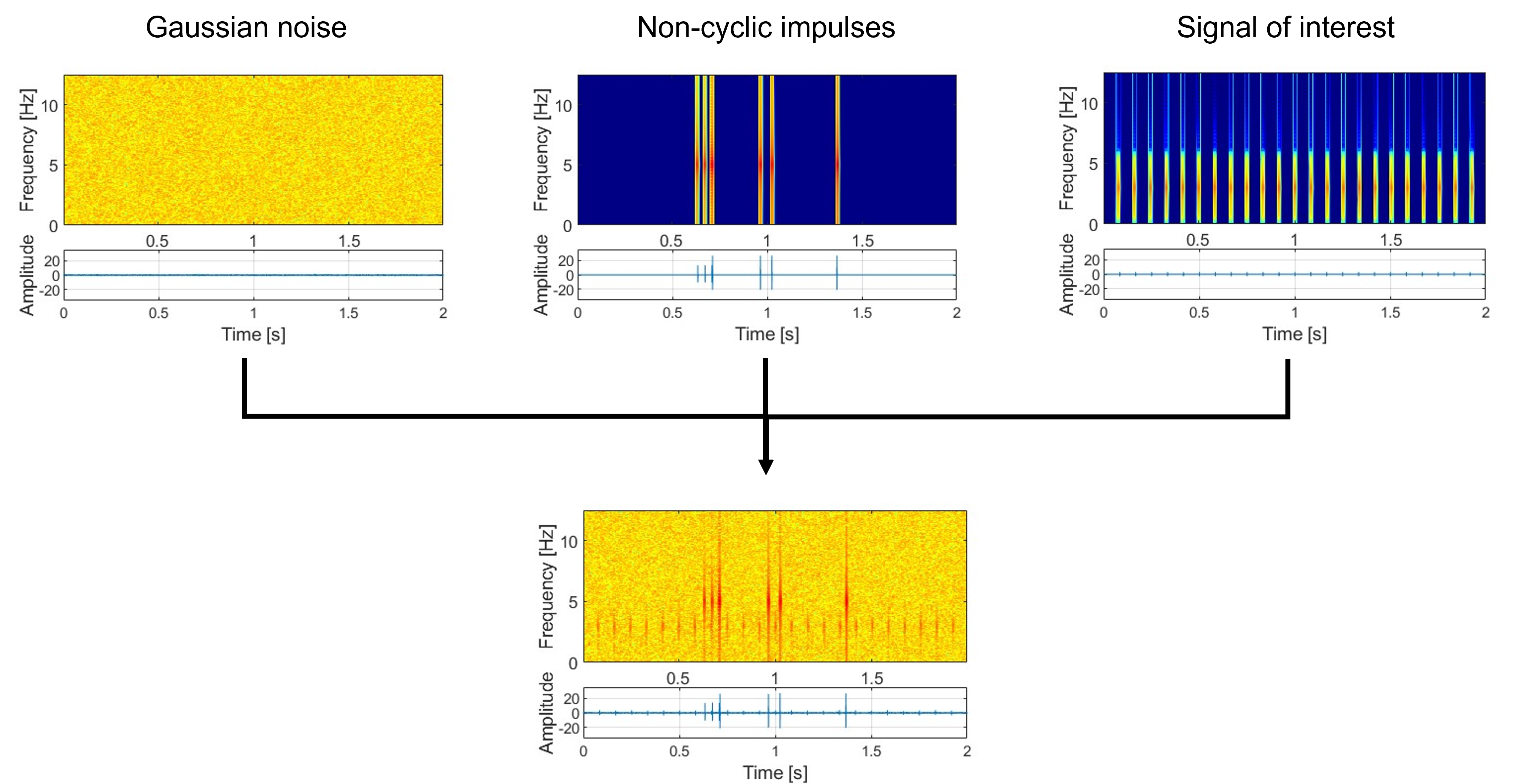}
\caption{Construction of the analyzed signal.} \label{comp}
\end{figure}

In this article, the authors used a simple and additive model of the signal.  This was constructed by assembling three elements according to the scheme shown in Figure \ref{comp}. These elements were Gaussian noise (with a mean of zero and a variance of $\sigma$), non-cyclic impulses, and the signal of interest (SOI).

The non-Gaussian noise, which is present in this scenario, is characterized by the multinomial distribution, see Section \ref{s:NC-pulses}, and corresponds to the non-cyclic impacts. The last discussed component is the cyclic impulsive signal, which represents the fault observed in the real-life signal, Section \ref{s:SOI}

\subsubsection{Non-cyclic impulses}\label{s:NC-pulses} 

The authors utilized the multinomial distribution to replicate the non-Gaussian behavior observed in the signal. It allows one to obtain the set of impulses which disturb the input signal. The density of this distribution is given by the formula \cite{alam1979estimation}:

\begin{equation}
\mathbb{P}(X_1=x_1, ..., X_N=x_N)=\frac{n!}{x_1!...x_N!}p_1^{x_1} \cdot...\cdot p_N^{x_N},
\end{equation} 
{where $\mathbb{X} = (X_1, X_2, ..., X_N)$ represents a random vector indexed by $N$ and $p_n$ denotes the probability of event $n$, with $p_n \in [0, 1]$ for $n \in (1, ..., N)$, and $\sum_{n=1}^N p_n = 1$.}.

In the paper, the authors decide to simulate {non-cyclic impulses} using a multinomial distribution with the number of possible events $N = 3${, that is:}

\begin{multline}
\mathbb{P}(X_1=x_1, X_2=x_2, X_3=x_3) = \\
= \frac{3!}{x_1!\cdot x_2!\cdot x_3!} p_1^{x_1}\cdot p_2^{x_2}\cdot p_3^{x_3} = \\
= \left\{ 
\begin{array}{lr} 
        x_1 = 0 &p_1^{x_1} = 1-p_2^{x_2}-p_3^{x_3}\vspace{4.5pt}\\
        x_2 = k_1 &p_2^{x_2}  \vspace{4.5pt}\\
        x_3 = k_2 &p_3^{x_3}. 
\end{array}\right.
\end{multline}

\subsubsection{Signal of interest (SOI)}\label{s:SOI}

The SOI can be characterized as a set of impulses distributed in the time domain with period $T$. It is translated into the fault frequency $f_{mod}=\frac{1}{T}$.
The single SOI impulse can be characterized as a decaying harmonic oscillation, which is described by the equation: \cite{wodecki2021influence}:

\begin{equation}\label{separate_imp}
h(t) = Amp \cdot sin(2\pi f_ct) \cdot e^{-dt},
\end{equation} 
where the amplitude of the impulse is represented by \textit{Amp}, time is denoted by \textit{t}, the carrier frequency is denoted as $f_c$, and the decay factor is represented by \textit{d}.

\subsection{Short-time Fourier transform \label{s:stft}}

The STFT is time-frequency decomposition defined as \cite{allen1977short}:  

\begin{equation}
STFT(f,t)=\sum_{l = 0}^{L-1} y(t + l) w(l) e^{\frac{2 \pi t l j}{\mathcal{F}}},
\end{equation}
where $y$ is input signal, $w(\cdot)$ is the shifting window,  $l = 1,2,...,L-1$ is window samples operator,  $t = 1,2,...,\mathcal{T}-1$ is the time point, $0 \leq f \leq \mathcal{F}-1$ is the frequency bin and \textit{j} is the imaginary operator.

The acoustic signal can be represented using a spectrogram as an absolute value of STFT at every time-frequency point:

\begin{equation}
\textbf{S}(f,t) = |STFT(f,t)|. 
\end{equation} 

A spectrogram, as a time-frequency representation of the signal, is a 2D matrix. It allows to divide signal through time into individual spectra changing over time, calculated from signal segments of length determined by the window size.
This property may be used to separate segments that contain wide-band, non-cyclic impulses. 


\subsection{Density-based spatial clustering of applications with noise {(DBSCAN)} } \label{s:dbscan}

DBSCAN is non-parametric algorithm used for clustering points with multiple nearby neighbors \cite{ester1996density}. The algorithm assigns every point to one of three states:

\begin{itemize}
\item \textbf{Core points}, where in defined distance $\epsilon$ at least $MinPts$ points are present
\item \textbf{Boundary points} reachable from core points, for a boundary less than $MinPts$ points are reachable within the distance $\epsilon$. A boundary point is also a part of a cluster.
\item \textbf{Outliers}, for which no other point is reachable within the distance $\epsilon$.
\end{itemize}

\begin{figure}[ht!]
\includegraphics[width=0.44\textwidth, angle=0]{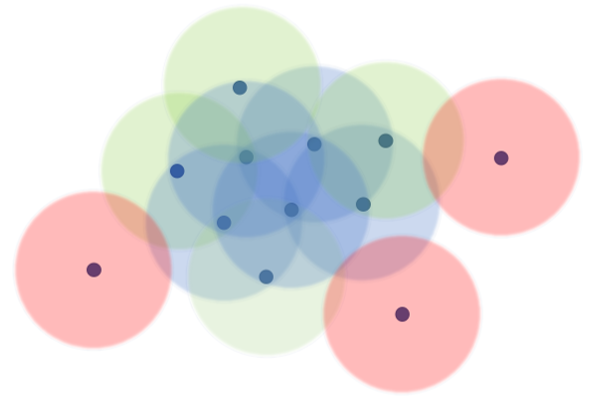}
\centering
\caption{DBSCAN schema for \textit{MinPts} = 4, blue - core, green - boundary, red - noise.} \label{fig1}
\end{figure}   
\unskip

%

\subsubsection{Parameters estimation \label{s:est}} 

\begin{itemize}
\item $\epsilon$: is the maximum distance for point adjacency

{To estimate the optimal {$\epsilon$}, the pairwise distances between all elements (spectral vectors of the spectrogram) are calculated first. In this study, the square Euclidean metric is used. The sought {$\epsilon$} value is given by the knee of the vector of distances sorted in descending order.}
It is a good estimator of {$\epsilon$} because larger values exceed within-class distances.

\item $MinPts$ is information about the minimum number of neighbours closer than {$\epsilon$} required to classify an element into the core of the class. 

\end{itemize}

\subsection{ The SOI component identification}

This procedure is divided into two steps: 
\begin{enumerate}
\item The DBSCAN algorithm applied to the spectrogram matrix enables the separation of the non-Gaussian impulsive component from the rest of the signal. The first iteration identifies the non-Gaussian impulsive component by density-based clustering. This process yields a vector of values containing $-1$ and $1$, where $-1$ indicates belonging to Class 1 and $1$ signifies membership in the core class (used for the second classification).  
\item The second execution of the same procedure separates the SOI from the remaining background noise. The final result is partial matrices of each of the signal components $\mathbf{Z}_1, \mathbf{Z}_2,\mathbf{Z}_3$. The spectrogram can be rewritten as follows:
\begin{equation}
    \textbf{S} = \textbf{Z}_1+\textbf{Z}_2+\textbf{Z}_3,
\end{equation}
where $\mathbf{Z}_1$ corresponds to non-cyclic impulses (Class 1), $\mathbf{Z}_2$ to signals of interest (Class 2) and $\mathbf{Z}_3$ to the background noise (Class 3). In this context, the operation of addition means that if a given individual spectrum (for example $Z_1(:,t)$) contains non-zero values, then $Z_2(:,t)$ and $Z_3(:,t)$ contains only zeros.
\end{enumerate}


\subsection{Signal reconstruction \label{s:mgle}}


According to the method presented by Wodecki et al. \cite{wodecki2020separation}, it is possible to invert the partial short-time Fourier transform of each component to reconstruct the time series of the components separately. However, it requires a complex STFT representation as input. 
To perform phase estimation, the authors suggest using the Griffin-Lim algorithm, first introduced in 1984 by Griffin and Lim \cite{griffin1984signal}. This algorithm is a variant of the double-projection algorithm.

The problem can be divided into two steps. First, for each of the three obtained partial matrices $\textbf{Z}_i$ for $i=\{1,2,3\}$ the goal is to derive a complete complex representation of the STFT. This representation should accurately characterize the signal with a magnitude that is very similar to the original signal. The obtained complex matrix that includes estimated phase is then used in the inverse short-time Fourier transform (iSTFT) to reconstruct the signal in the time domain.

\subsection{Squared envelope spectrum based indicator}

The cyclic impulsive signal in the time domain associated with the diagnosed damage can be represented as a family of spectral components (fault frequency and harmonics) in the envelope spectrum. 
To verify the results of the presented method, the ENVSI proposed by Hebda-Sobkowicz \cite{hebda2020informative} was used. 

It is an indicator based on the squared amplitude of the information signal, i.e., $AIS = SES(f_{mod}(1,\dots,M_1))$ and the square of squared envelope spectrum (SES):
\begin{equation}\label{eq:envsi}
    ENVSI=\frac{\sum_{i=1}^{M_1}AIS^2(i)}{\sum_{k=1}^{M_2}SES^2(k)},
\end{equation}
where $M_1$ is the number of harmonics and $M_2$ are {number of spectral bins} used for the analysis (trimmed to the last component of considered harmonics).

\subsection{The fault frequency identification}

A simple peak detection method was used to identify informative (fault-related) components. Peaks of interest were selected based on their prominence and the differences between the positions of the identified local maxima were calculated. The median of this series of differences is used as the desired result.

\subsection{Selectors}
\subsubsection{Spectral Kurtosis}
The kurtosis statistic is an important measure used in probability and statistical theory to analyze the shape of a probability distribution. It provides information about the tails of the distribution. When the signal has a Gaussian distribution, the kurtosis statistic is equal to 0, indicating that the tails of the distribution are similar to those of a normal distribution. The Spectral Kurtosis, initially proposed by Antoni and Randall (\cite{antoni2007cyclic,antoni2006spectral}), is widely recognized as a powerful and popular method to identify an informative frequency band (IFB), especially in vibration signal analysis. The fundamental process involves calculating the estimator of the kurtosis value for every frequency bin in the spectrogram $\mathbf{S}=[\mathbf{s}_{f_1},...,\mathbf{s}_{f_\mathcal{F}}]$.

\begin{eqnarray}\label{eq:kurtosis}
\widehat{\textrm{SK}}(\mathbf{s}_{f_k})=\frac{\frac{1}{\mathcal{T}}\sum_{i=1}^{\mathcal{T}}\left(s_{f_ki}-\overline{\mathbf{s}}_{f_k}\right)^4}{\left(\frac{1}{\mathcal{T}}\sum_{i=1}^{\mathcal{T}}\left(s_{f_ki}-\overline{\mathbf{s}}_{f_k}\right)^2\right)^2} -3,
\end{eqnarray} 

where $\overline{\mathbf{s}}_{f_k}$ denotes the sample mean and $\mathcal{T}$ represents the sample length of single time-domain vector corresponding to the given frequency bin.

An indicator of impulsiveness across the signal spectrum can be derived to identify the frequency bands that contain the most prominent information regarding impulsive behavior related to damage. This indicator can be utilized as a filter to extract the impulsive components from the signal.

\subsubsection{Alpha-selector}

 The second selector is based on the stability index $\alpha$ of the $\alpha$-stable distribution, which also utilizes the impulsive characteristic of SOI. 
The cumulative distribution function (CDF) and the probability distribution function (PDF) of $\alpha$-stable distributions are generally not expressed in closed forms. The random variable $X$ is considered $\alpha$-stable if its characteristic function is defined as \cite{Taqqu}:

\begin{eqnarray}\label{characteristic_function}
		\varphi_X(t) =
		\left\{
		\begin{array}{ll} 
		\exp\left\{-\sigma^{\alpha}|t|^{\alpha}\left\{1-j\beta \mathrm{sign}(t)\tan\left(\pi\alpha/2\right)\right\}+j\mu t\right\}& {\textnormal{for}\ } \alpha\neq 1,\\
		&\\
		\exp\left\{-\sigma|t|\{1+j\beta \mathrm{sign}(t)\frac{2}{\pi}\log(|t|)\}+j\mu t\right\}& {\textnormal{for}\ } \alpha=1,
		\end{array}
		\right.
	\end{eqnarray}
where $j$ is the imaginary operator, $\alpha \in (0,2]$ represents the stability index, $\beta$ signifies the skewness, $\sigma$ denotes the scale parameter and $\mu$ indicates the shift parameter.
When the parameter of an $\alpha$-stable distribution is considered, it is important to note that when $\alpha=2$, the distribution simplifies to the Gaussian distribution. In this case, the Gaussian distribution is characterized by a variance of $2\sigma^2$ and a mean of $\mu$. As the parameter $\alpha$ decreases, the tails of the distribution exhibit a heavier tail behavior. The Alpha-selector is defined with the above properties in consideration \cite{zak2015application}. Similarly, as in the SK, the estimator of the stability index $\alpha$ is calculated for every frequency bin in the spectrogram $\mathbf{S}=[\mathbf{s}_{f_1},...,\mathbf{s}_{f_\mathcal{F}}]$.

\begin{equation}\label{eq:alpha-sel}
    \widehat{\textrm{Alpha}}(\mathbf{s}_{f_k})=2-\hat{\alpha} (\mathbf{s}_{f_k}).
\end{equation}

A selector constructed according to Equation (\ref{eq:alpha-sel}) will have values close to 0 for frequency bins that approximate a Gaussian distribution. As the stability index decreases, the selector values at that location increase. The Alpha selector has proven to be a valuable tool in the process of identifying IFB in bearing fault analysis, particularly with non-Gaussian impulsive background noise \cite{HebdziaSobkowicz2020b}.

\subsubsection{Conditional variance selector}
The third selector chosen for the comparison is based on the conditional variance. The concept of conditional variance is related to the statistical principle known as the 20/60/20 Rule. This rule suggests that when a population is divided into three groups based on a specific criterion (e.g., 20\% of the smallest values, 60\% of the middle values, and 20\% of the largest values), it often indicates some balance within the data. If the data comes from a Gaussian distribution, then the conditional variance in these groups will be approximately the same. Note, that varying partitioning sets could be taken into consideration \cite{jelito2021new}. In \cite{HebdziaSobkowicz2020b}, the selector is based on dividing the data into seven parts. In this case the unique quantile ratio, given the equality of variance, can be defined as follows:
\begin{equation}
    0.004/0.058/0.246/0.384/0.246/0.058/0.004
\end{equation}

The estimators of the seven quantile partitioning for each frequency bin in the spectrogram are defined as follows:
\begin{equation}\label{AA}
    \begin{aligned}
\hat{A}_1 &:=(-\infty,~\hat{\mathbf{s}}_{f_k,0.004}],\\
\hat{A}_2 &:=(\hat{\mathbf{s}}_{f_k,0.004}, ~\hat{\mathbf{s}}_{f_k,0.062}],\\
\hat{A}_3 &:=(\hat{\mathbf{s}}_{f_k,0.062},~\hat{\mathbf{s}}_{f_k,0.308}],\\
\hat{A}_4 &:=(\hat{\mathbf{s}}_{f_k,0.308},~\hat{\mathbf{s}}_{f_k,0.692}],\\
\hat{A}_5 &:=(\hat{\mathbf{s}}_{f_k,0.692},~\hat{\mathbf{s}}_{f_k,0.938}],\\
\hat{A}_6 &:=(\hat{\mathbf{s}}_{f_k,0.938},~\hat{\mathbf{s}}_{f_k,0.996}]\\
\hat{A}_7 &:=(\hat{\mathbf{s}}_{f_k,0.996},~\infty),
    \end{aligned}
\end{equation}
where  
$\hat{\mathbf{s}}_{f_k,q}$ is the empirical quantile of order $q$ calculated for vector $\mathbf{s}_{f_k}$. The estimator of the test statistic considered as a selector is defined as:

\begin{eqnarray}\label{stat_N2}
\widehat{CV}(\mathbf{s}_{f_k})
:=\left(\frac{\hat{\sigma}^2_{A_3}-\hat{\sigma}^2_{A_4}}{\hat{\sigma}}+\frac{\hat{\sigma}^2_{A_5}-\hat{\sigma}^2_{A_4}}{\hat{\sigma}}\right)^2\sqrt{\mathcal{T}},
\end{eqnarray}
where $\hat{\sigma}_{A_i}$ represents the estimator of the standard deviation in the set $A_i$.

The selector obtained according to this procedure is called CV-selector. 







\section{Results}\label{sec3}

In order to test the {efficiency} of the proposed {methodology}, a multiple signal simulations were carried out. Then, the steps proposed in the methodology were implemented, aimed at separating the response signal elements and trying to recover the frequency of interference resulting from the signal of interest component. Finally, the real data with the faulty bearing used in the industry was considered to confirm the effectiveness of the proposed methodology. 
 
\subsection{Simulations}\label{subsec3.1}
This section presents the results obtained for the simulated data. The synthetic signals were simulated with a length of 2 seconds (50,000 samples), sampled at 25 kHz. As presented in Figure \ref{comp}, the signal is composed of three components. They are simulated separately and finally added together:
\begin{itemize}
    \item \textbf{Gaussian noise} - It is assumed that the analyzed signal is centred around zero ($\mu=0$). In the analysis, six different levels of the scale parameter $\sigma$ were considered. It increases from $0.6$ to $1.6$ with step $0.2$. The variability of the $\sigma$ parameter describes how much the SOI is hidden in the noise.
    \item \textbf{non-Gaussian noise} - The non-cyclic impulses were generated according to the following procedure. First, the amplitudes $k_1=15$ and $k_2=30$ were drawn from a trinomial distribution with a given probability. The probability of occurrence of a given event was selected so that the number of events in a sequence of 1s could be determined (denoted $\gamma$). In the presented analysis, this parameter was defined at 7 levels, increasing from 1 to 7. This implies that, for example, for $\gamma=1$, one impulse of amplitude $k_1$ and one of amplitude $k_2$ can be expected. Similarly to SOI, each impulse can be characterized by a decaying harmonic oscillation with the following parameters: $k_1$ and $k_2$ are the amplitudes, carrier frequency $f_c = 5000$ Hz, parameter $d = 1800$.
    \item \textbf{SOI}  - the cyclic impulsive component was generated according to Eq. \ref{separate_imp} for the same parameter for each scenario, namely: amplitude $Amp = 1$, carrier frequency $f_c = 2500$, decay parameter $d = 1800$ and modulating frequency {$f_{mod} = \frac{1}{T}=30.7$ Hz.}
\end{itemize}

\begin{figure}[ht!]
\centering
\includegraphics[width=0.75\textwidth, angle=0]{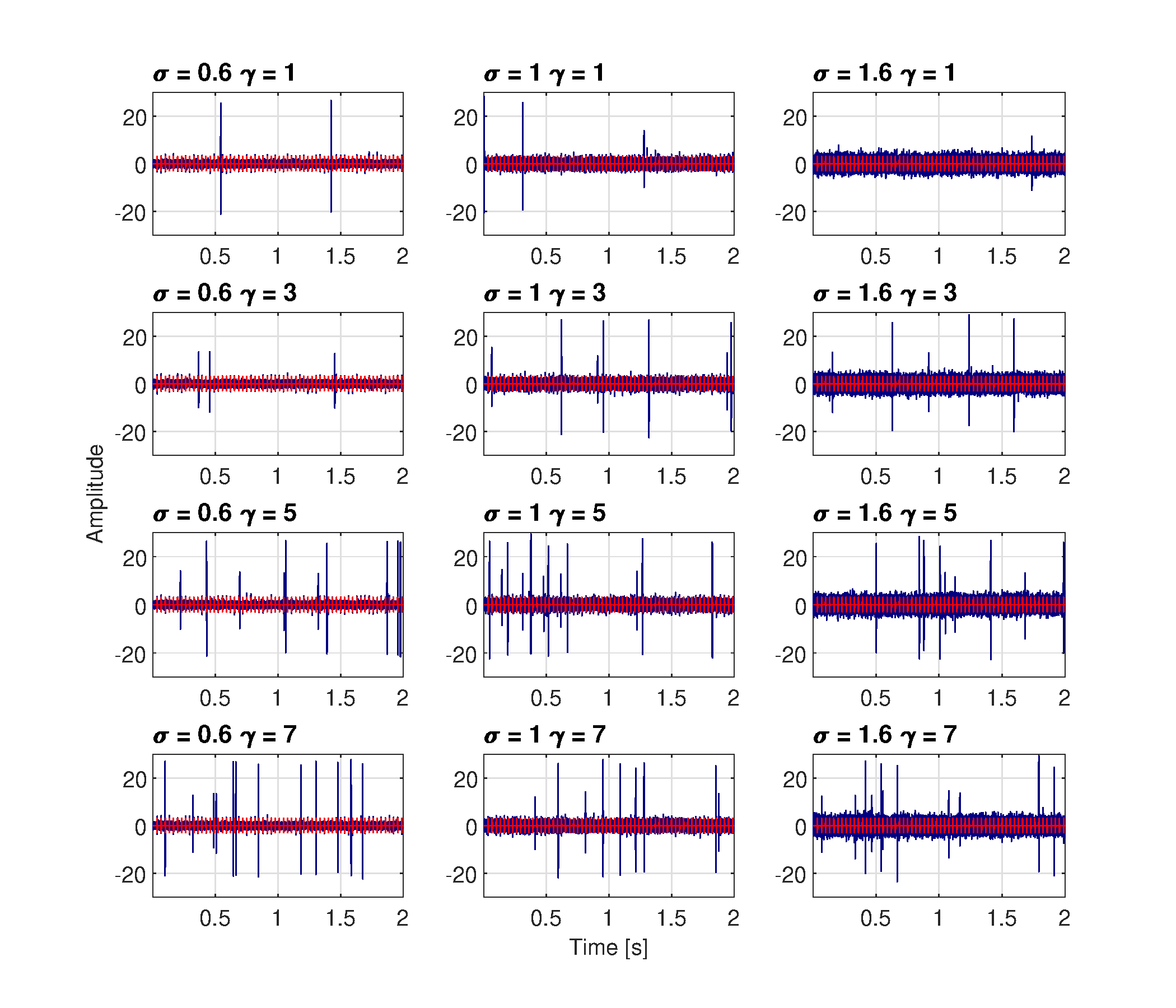}
\caption{Exemplary time series of simulated signal.}\label{f: time}
\end{figure}

In Figure \ref{f: time} the exemplary time series for the selected parameters $\sigma$ and $\gamma$ are presented.
In the first row, $\gamma = 1$, which means that one non-cyclic impulse per second can be expected. 
As the $\gamma$ parameter increases, the number of non-cyclic disturbances is higher. 
Similarly, the effect of the $\sigma$ parameter on the behavior of the time series is presented. In order to better visualize the behavior of the SOI relative to the rest of the signal, it is additionally drawn in red. It is easier to see how the {$\sigma$} parameter affects the visibility of SOI. In the first column $\sigma=0.6$, the cyclic impulses associated with the fault are visible and extend above the noise.
In the second column $\sigma=1$, a limiting situation is presented - cyclic impulses have the same amplitude as noise. 
In the last column $\sigma=1.6$, the SOI is completely hidden in the noise.

\begin{figure}[ht!]
\centering
\includegraphics[width=0.75\textwidth, angle=0]{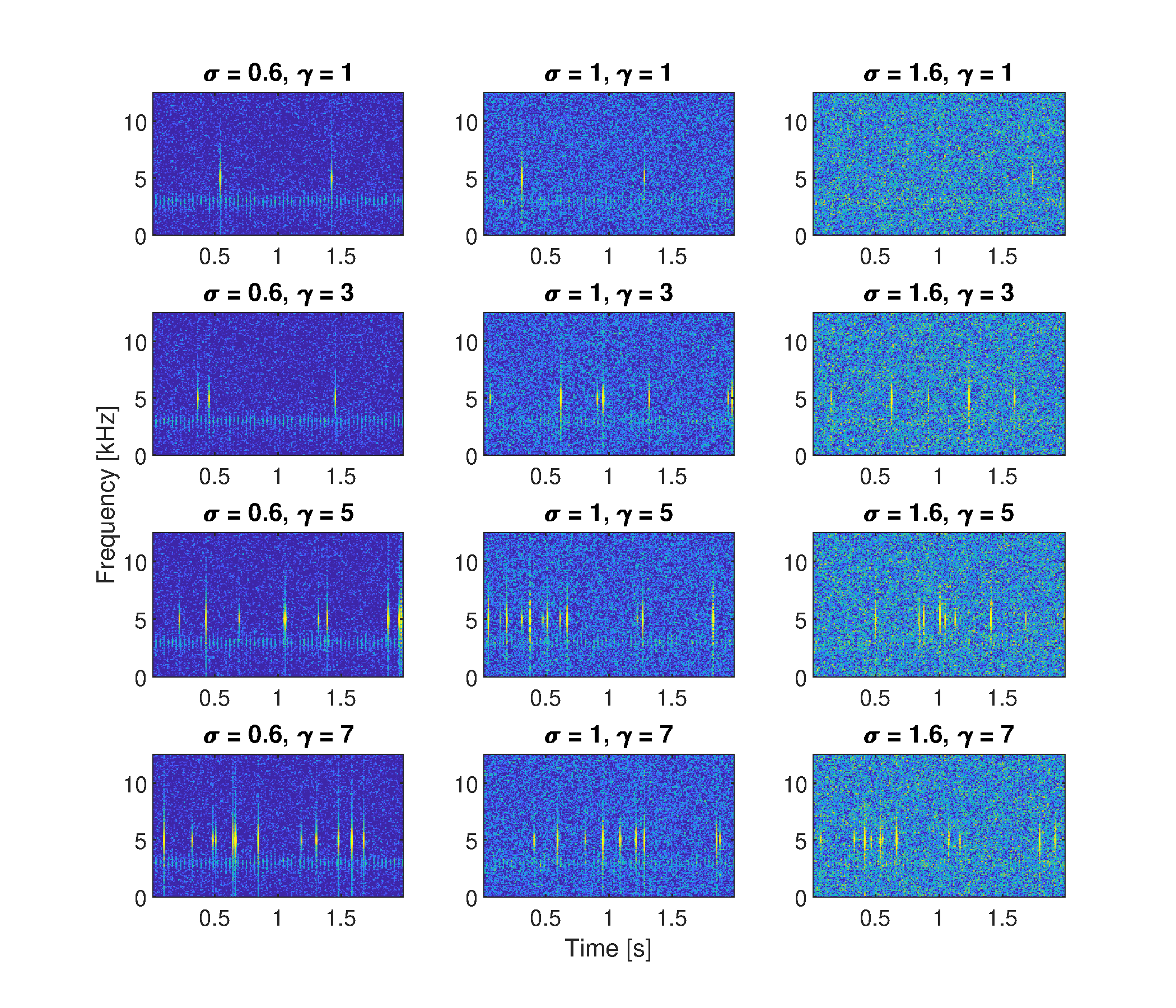}
\caption{Spectrograms of the exemplary simulated signals for fault frequency $f_{mod} = 30.7$ Hz.}\label{f:spectro}
\end{figure}  

 In the next step, the time-frequency representations for each signal were calculated with the following parameter: length of the window is 256, nfft is 512 and the overlap is 85\% of a window. In Figure \ref{f:spectro} the spectrograms corresponding to the time series from Figure \ref{f: time} are presented.  As the parameter $\gamma$ increases, the number of non-cyclic impulses increases. For all observed cases, regardless of the noise level, they are visible on the spectrograms.  When the $\sigma = 0.6$ both cyclic and non-cyclic impulses are clearly visible. The situation changes in the case of cyclic impulses - their visibility decreases as the sigma parameter increases.

\begin{figure}[ht!]
\centering
\begin{subfigure}{0.49\textwidth}
    \includegraphics[width=\textwidth, angle=0]{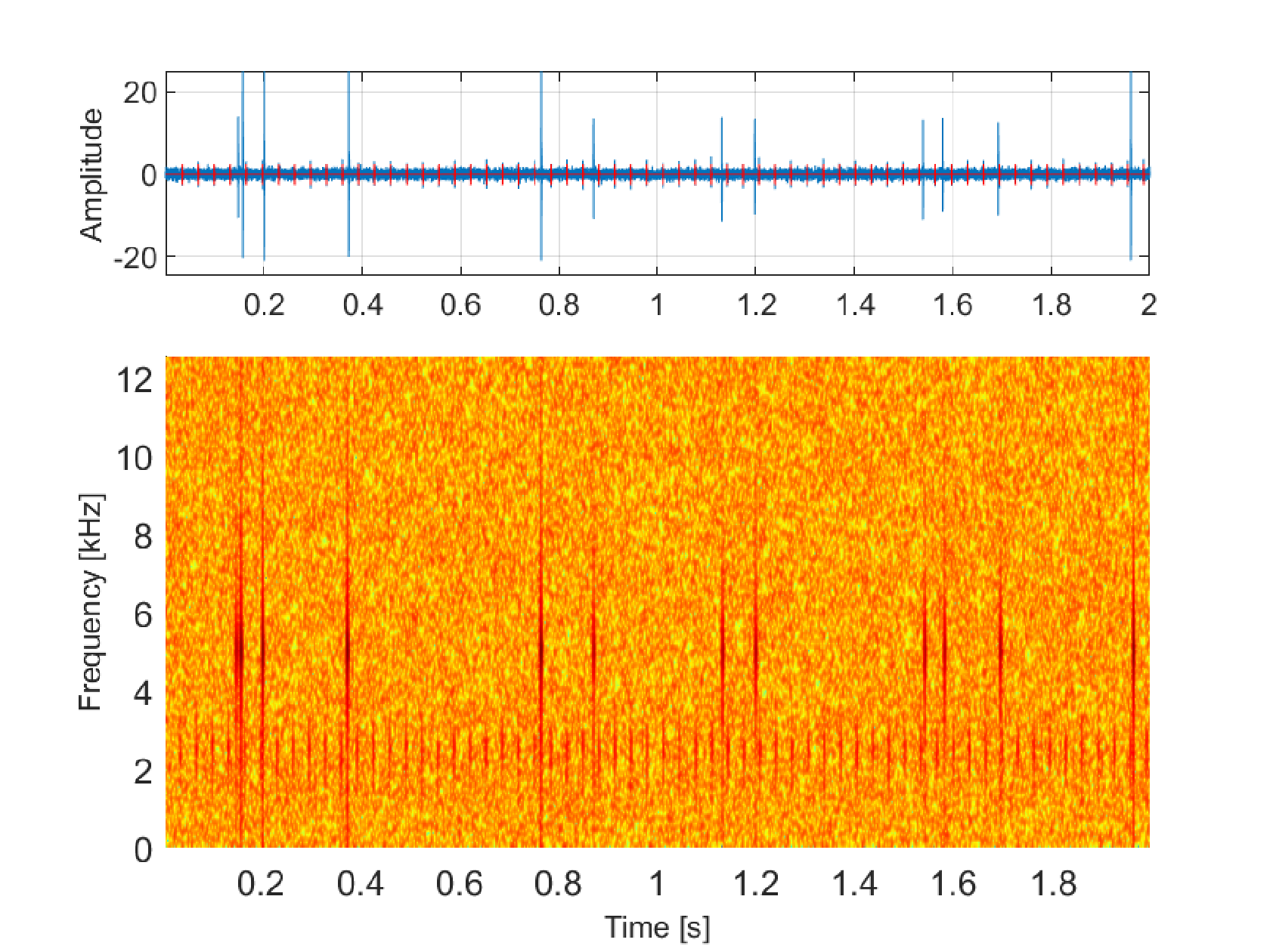}
    \caption{With damage.}
	\label{fig:subfigA}
\end{subfigure}
\begin{subfigure}{0.49\textwidth}
    \includegraphics[width=\textwidth, angle=0]{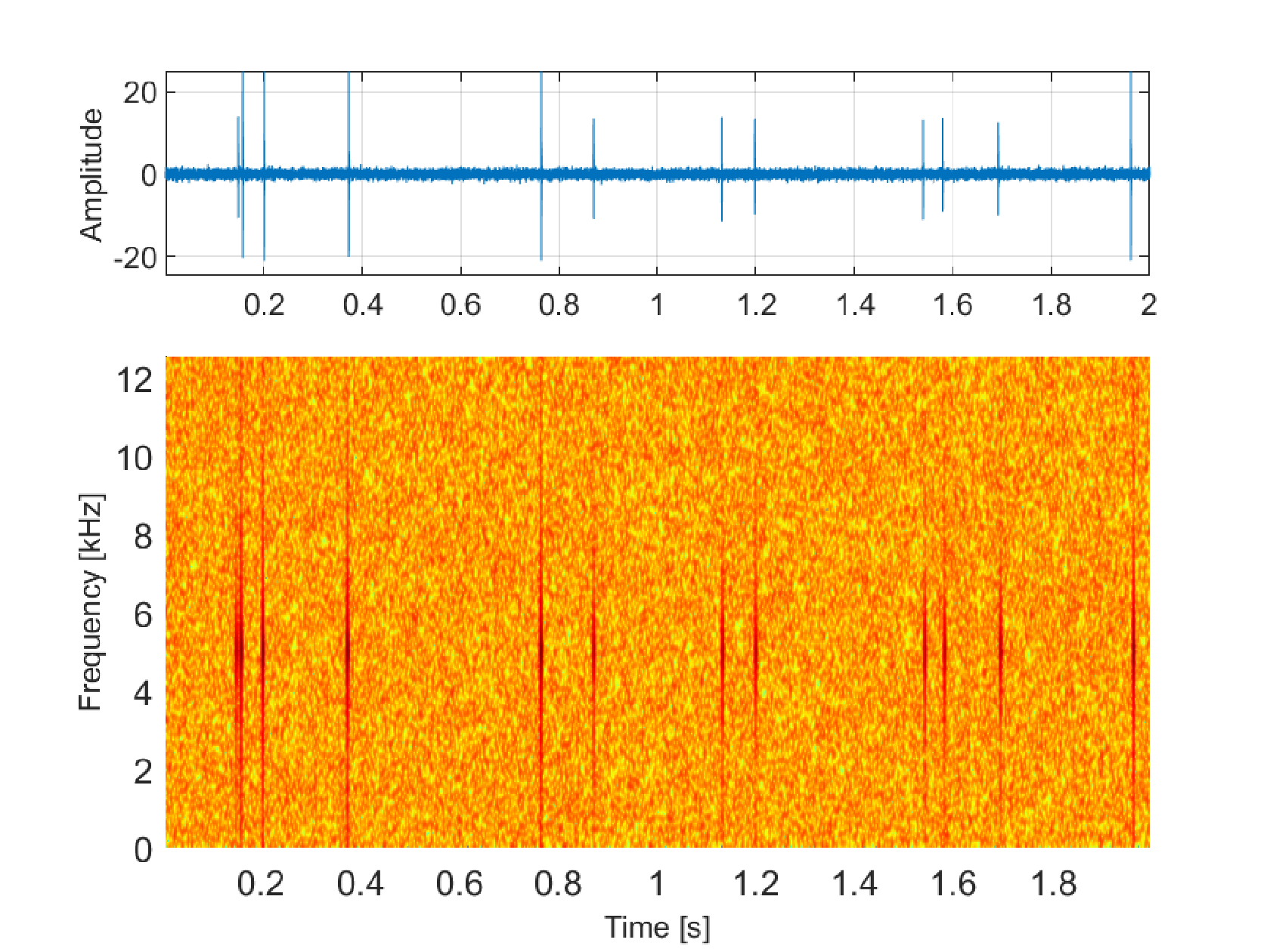}
    \caption{Without damage.}
	\label{fig:subfigB}
\end{subfigure}

\caption{Time series of simulated signal for $\sigma = 0.6$ and $\gamma=3$ with corresponding spectrogram. Left panel with added SOI component of $f_{mod} = 30.7$ Hz  and right panel without damage.}
\label{f:time1}
\end{figure}

For a single signal (with the parameters $\sigma = 0.6$ and $\gamma=3$), the individual steps of the algorithm are presented in detail. The time series of these selected signals are presented in Figure \ref{f:time1}.

The authors assumed that the spectrogram consists of a set of {spectral} vectors. For each of them, the distances between the remaining vectors were calculated using the quadratic Euclidean metric to apply DBSCAN clustering. After the appropriate parameters were selected, the DBSCAN procedure was used to separate non-cyclic signals.

\begin{figure}[ht!]
\centering
\begin{subfigure}{0.49\textwidth}
    \includegraphics[width=\textwidth, angle=0]{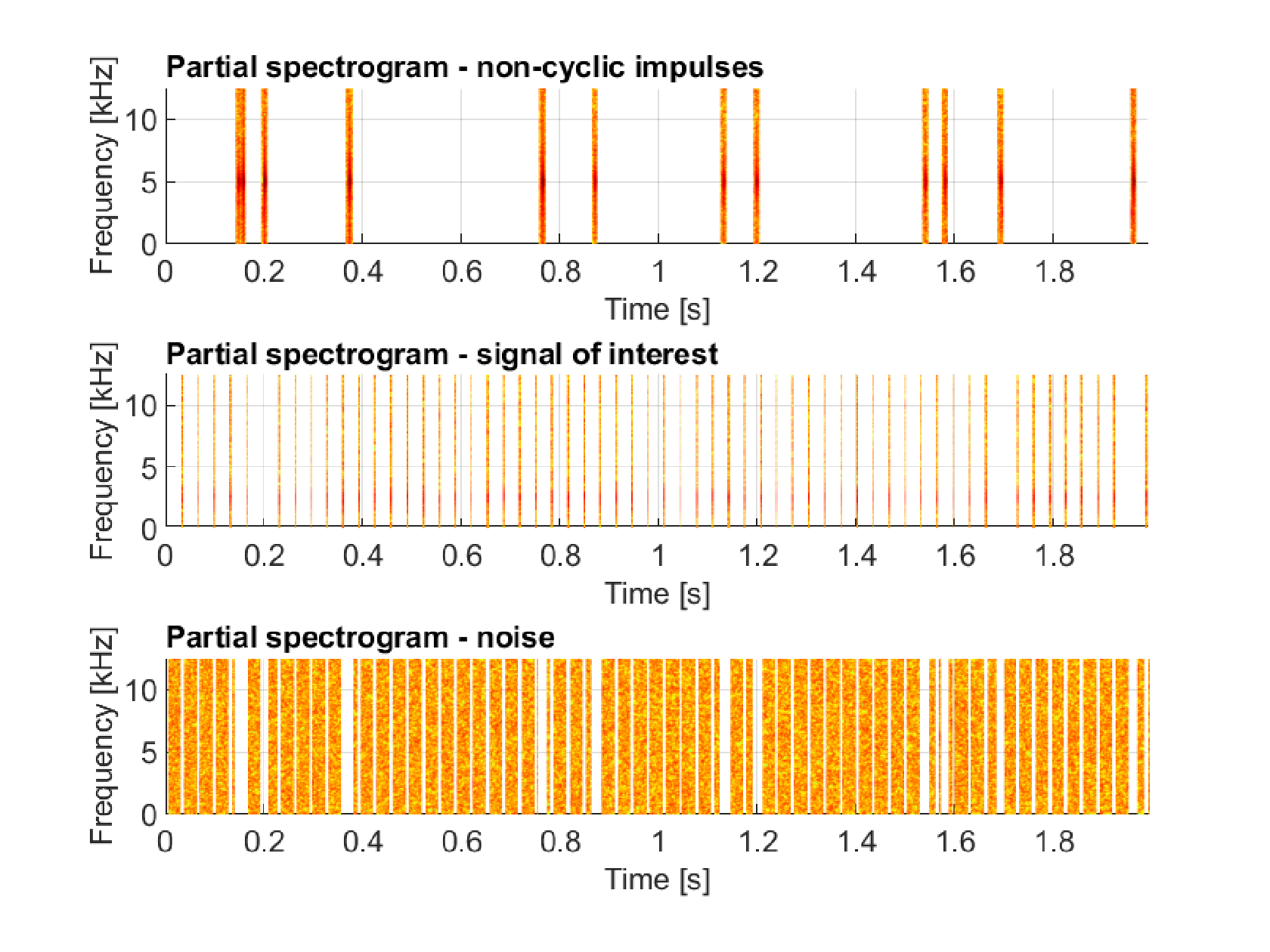}
    \caption{With damage.}
	\label{fig:classA}
\end{subfigure}
\begin{subfigure}{0.49\textwidth}
    \includegraphics[width=\textwidth, angle=0]{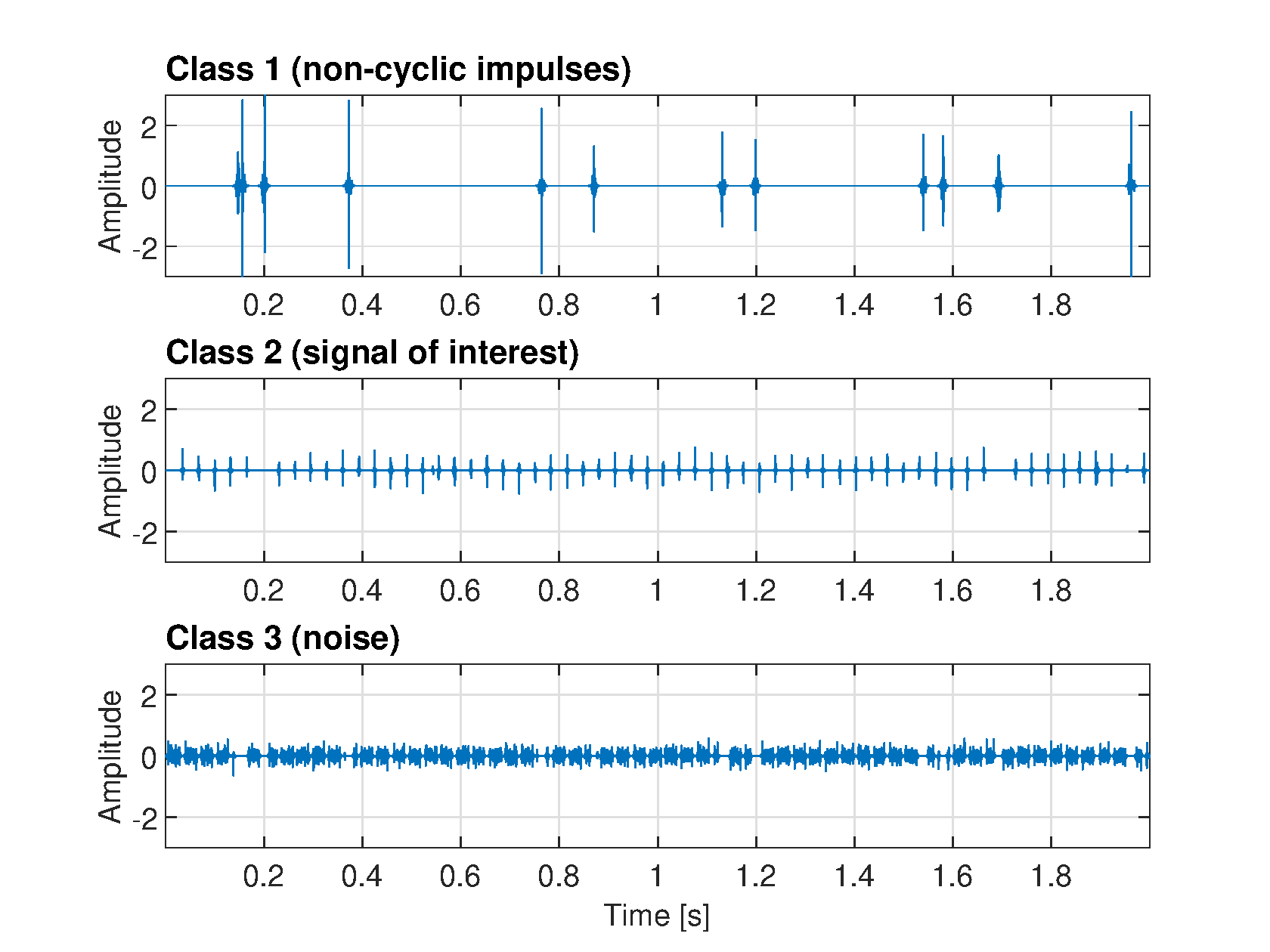}
    \caption{With damage.}
	\label{fig:timeA}
\end{subfigure}
\caption{Partial spectrograms divided by classes 1-3 with related time series for a signal with SOI component.}\label{f:specSOI}
\end{figure}

\begin{figure}[ht!]
\centering
\begin{subfigure}{0.49\textwidth}
    \includegraphics[width=\textwidth, angle=0]{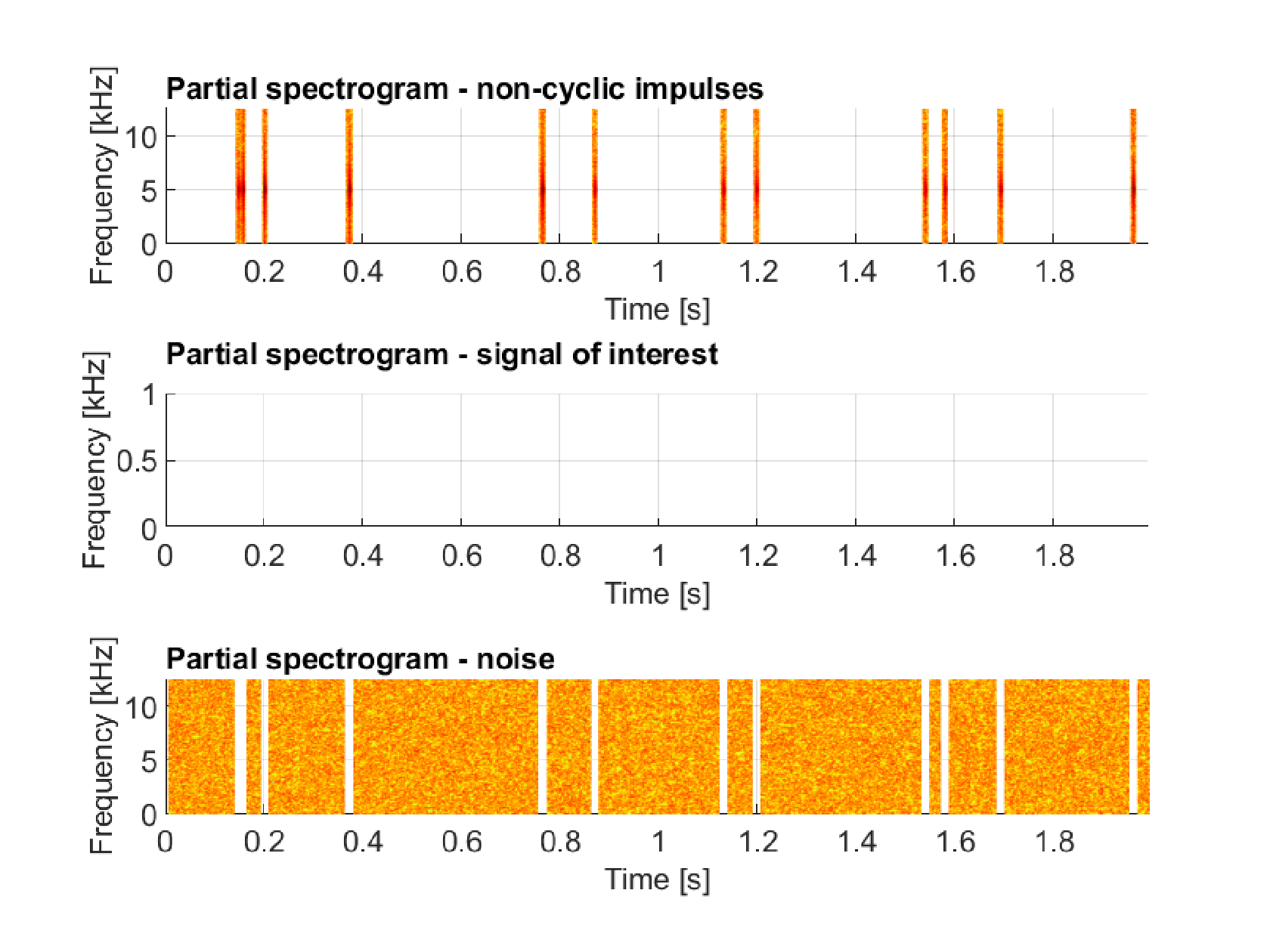}
    \caption{Without damage.}
	\label{fig:classB}
\end{subfigure}
\begin{subfigure}{0.49\textwidth}
    \includegraphics[width=\textwidth, angle=0]{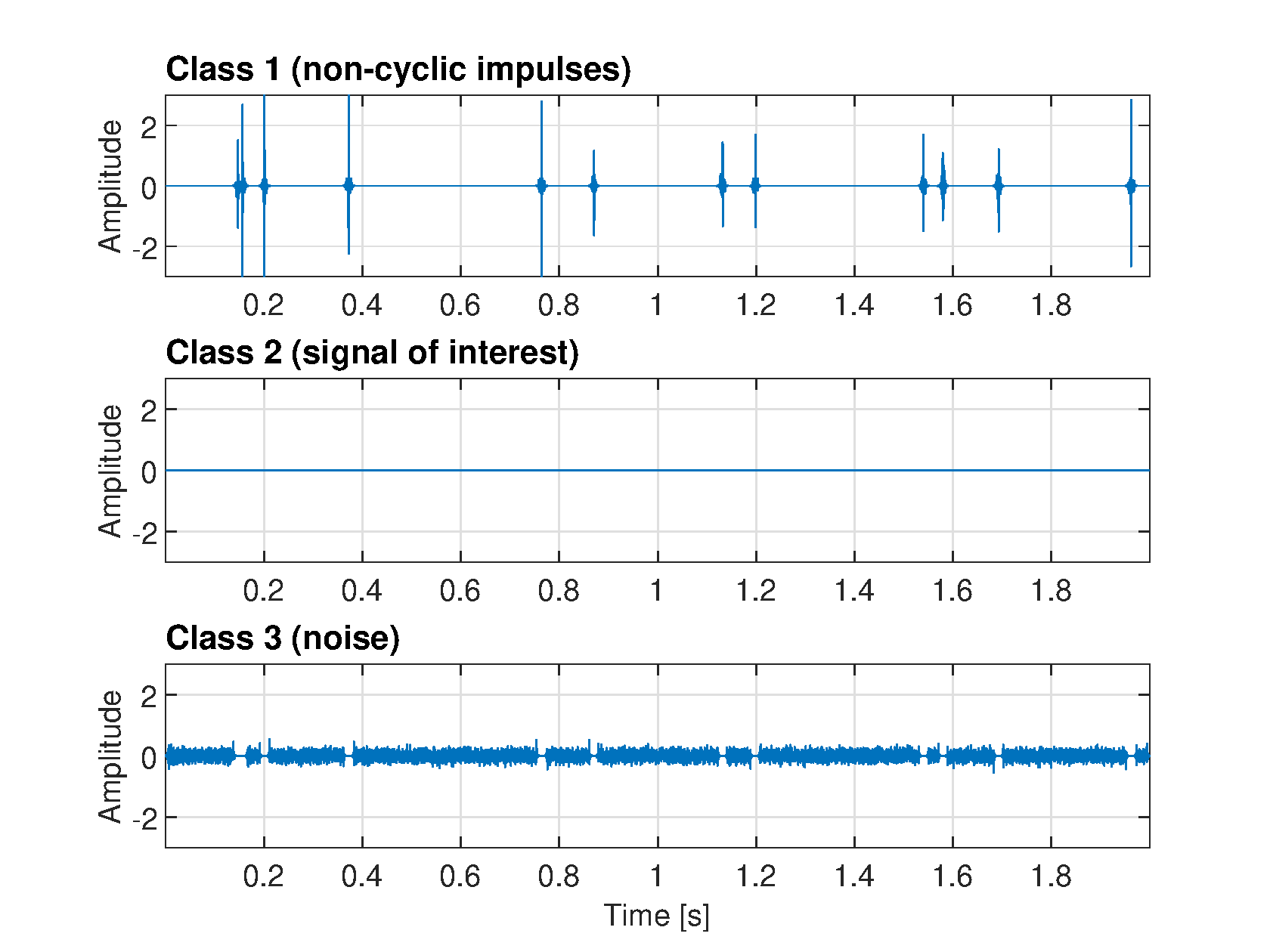}
    \caption{Without damage.}
	\label{fig:timeB}
\end{subfigure}
\caption{Partial spectrograms divided by classes 1-3 with related time series for a signal without SOI component.}\label{f:specNO}
\end{figure}

Figures \ref{fig:classA} and \ref{fig:classB} shows that the separation worked correctly for all non-cyclic {and cyclic} disturbances in the signal. The algorithm classified the input data into three classes.

In the first step, Class 1 contains non-classified elements, which (in the presented case) corresponds to the non-cyclic impulses. The remaining signal components (that is, Gaussian noise and SOI) were classified as Class 2.

To maintain the dimensions of the considered matrix and avoid changing the statistics, both classes were separated by filling the gaps with NaN-s. The next step is to separate the cyclic signals from the noise by repeating the DBSCAN procedure. Therefore, the parameters for the second clustering of the data should be recalculated.

After re-executing the proposed procedure, the class corresponding to the non-classified elements associated with cyclic disturbances (SOI) and the class containing vectors representing Gaussian noise were obtained.
The influence of distortions resulting from the spectrogram's specificity is also noticeable. 

For further analysis, it is necessary to recover the time series of the signal from the spectrogram. It is possible to use the Griffin-Lim algorithm, as was described in Section \ref{s:mgle}. Based on the partial spectrograms, three time series were recovered (see Figures \ref{fig:timeA} and \ref{fig:timeB}).


\begin{figure}[ht!]
\centering
\begin{subfigure}[b]{0.49\textwidth}
    \includegraphics[width=\textwidth, angle=0]{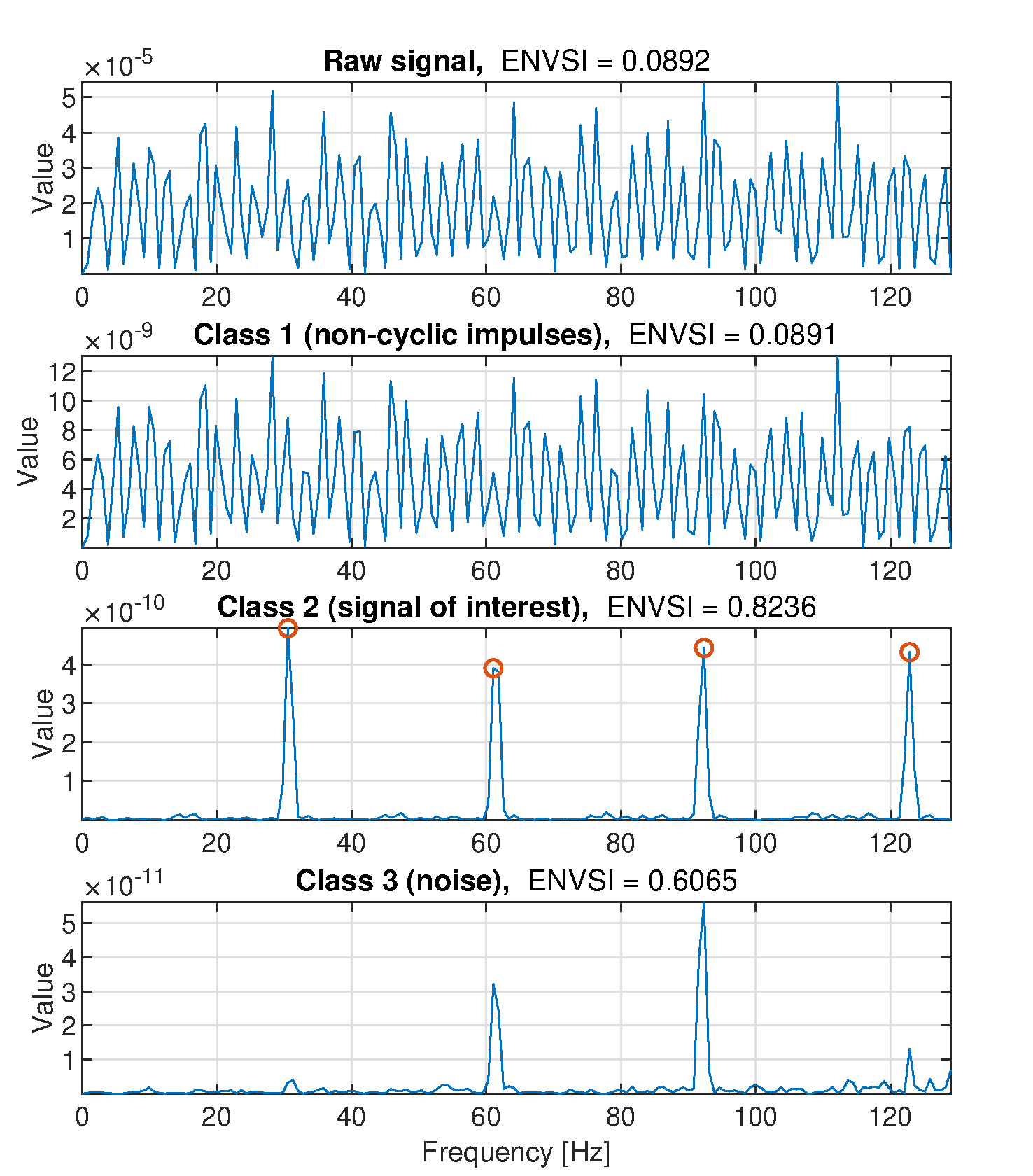}
    \caption{With damage.}\label{fig:envA}
\end{subfigure}
\begin{subfigure}[b]{0.49\textwidth}
    \includegraphics[width=\textwidth, angle=0]{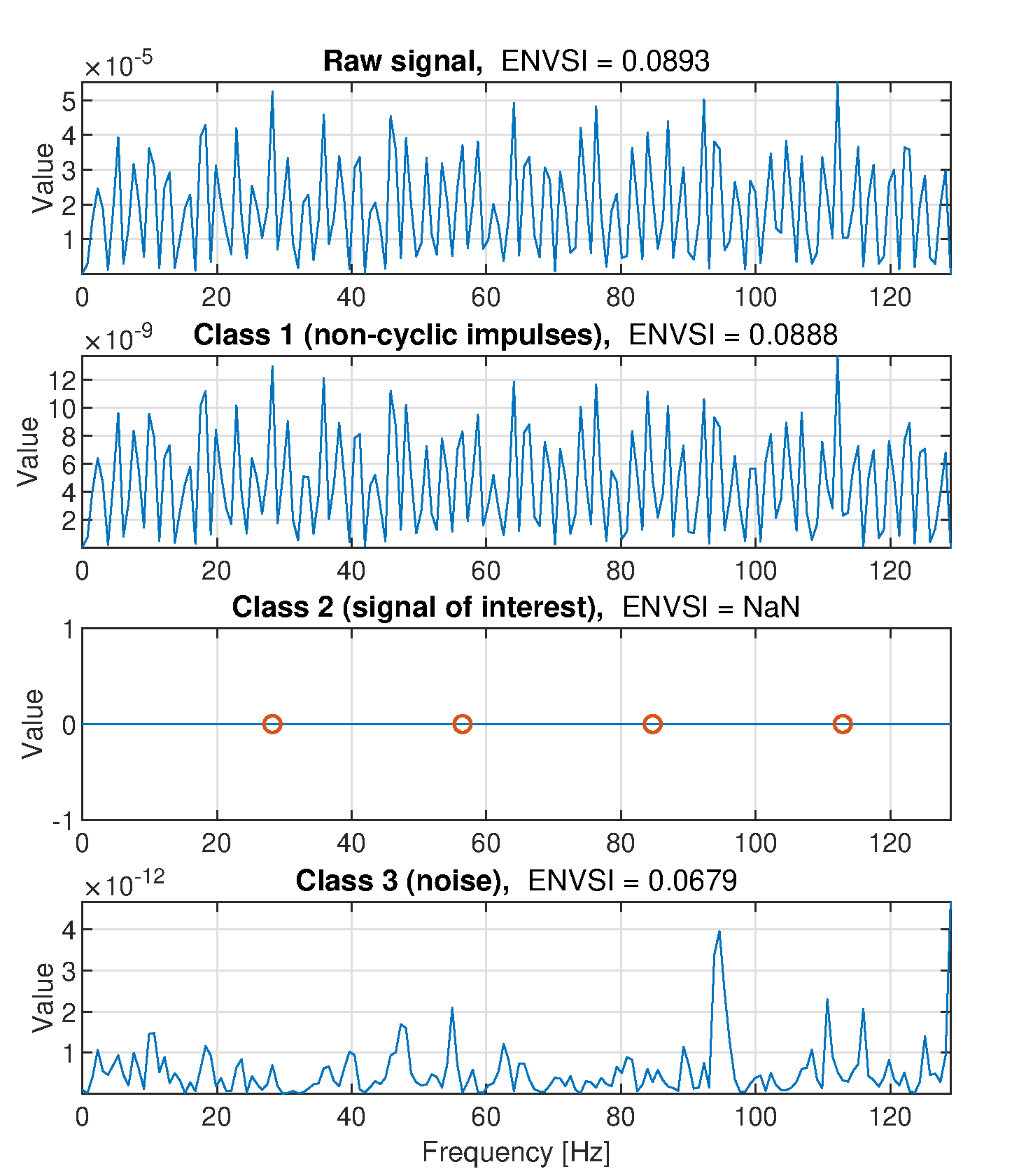}
    \caption{Without damage.}\label{fig:envB}
\end{subfigure}
\caption{Squared envelope spectrum of a damaged and healthy signal for classes 1-3.}\label{f:env}
\end{figure}  

Finally, the authors computed the envelope spectrum of the components obtained (see Figure \ref{f:env}). Based on these spectra and the information about the fault frequency (modulation frequency from the simulation), the effectiveness of the proposed algorithm was evaluated using ENVSI. For the presented simulation with the SOI component, the ENVSI value for the first class, corresponding to non-cyclic impulse disturbances, is the lowest at 0.09. The second class, which represents the SOI, has a high ENVSI value of 0.82. In addition, a high indicator value of 0.60 is observed in the third class. This phenomenon is due to the construction of the partial matrices - whenever values appear in class two, there are gaps in class three. Consequently, the spectrum from which ENVSI is calculated also manifests this frequency (see Figure \ref{fig:envA}).


\subsection{Real data}\label{subsec3.2}
 
In this Section, the results of the analysis of two acoustic signals are presented.

\subsubsection{Centrifugal pump}
\begin{figure}[ht!]
\centering
\includegraphics[width=.95\textwidth, angle=0]{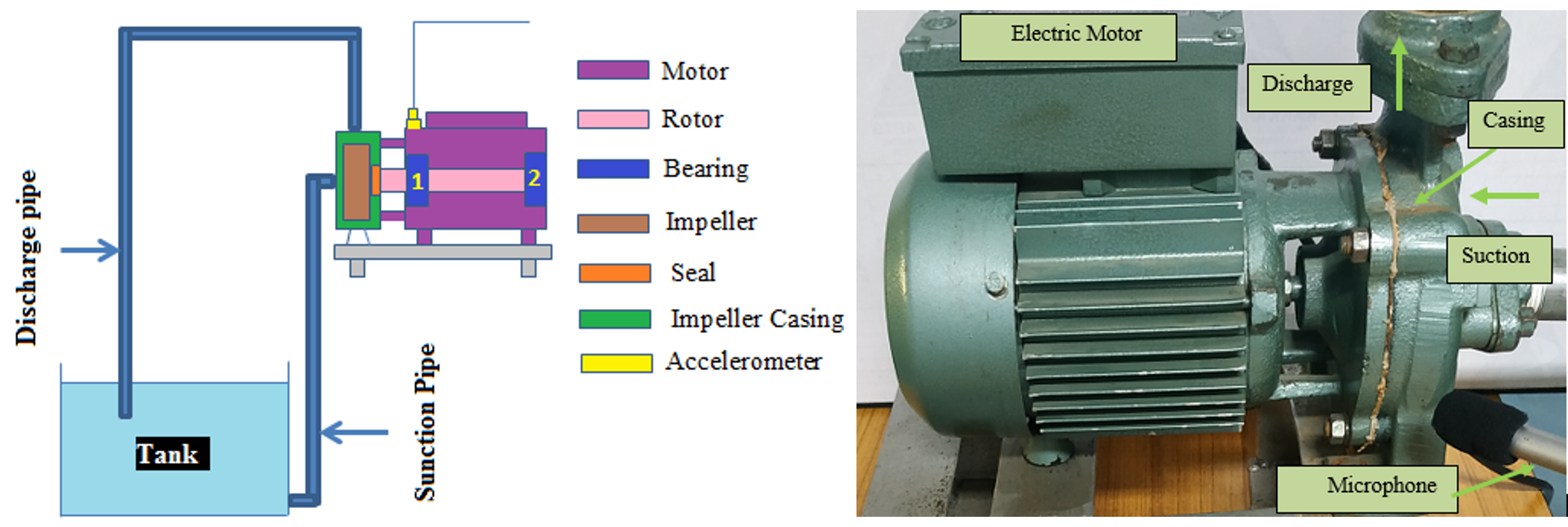}
\caption{The scheme (left panel) and the photo during the acoustic measurement session(right panel) of the test rig. }\label{f:ac_CP_test_rig}
\end{figure}

The experiment was conducted using a CRI Monoblock centrifugal pump (Model: ACM-0 (AF)) running at a speed of 43 Hz. The test rig scheme and the photo of it during the acoustic measurement session are shown in Figure \ref{f:ac_CP_test_rig}. The setup consists of a rotor supported by two bearings, designated Bearing 1 and Bearing 2. The bearing with the fault is called Bearing 1 (6203-ZZ bearing) and is located closer to the impeller. For a given rotational frequency, the damage to the outer ring has a fault frequency of 127 Hz.  The impeller overhangs the rotor shaft and is housed in the impeller casing. It has 3 rotating impeller vanes that draw in fluid axially through the impeller eye, imparting kinetic energy to the fluid, and causing the fluid to flow outward radially through the casing. 

The device used to collect acoustic data is the NI-USB-4431, as shown in Fig. 4. It is a portable 24-bit, 4-channel analog I/O device designed for gathering vibration data. The NI-USB-4431 includes signal conditioning modules with a low-pass Bessel filter to reduce aliasing. The acoustic data, used in this paper, was sampled at a frequency of 70 kHz and recorded using a microphone (ECM8000).

\begin{figure}[ht!]
\centering
\includegraphics[width=.8\textwidth, angle=0]{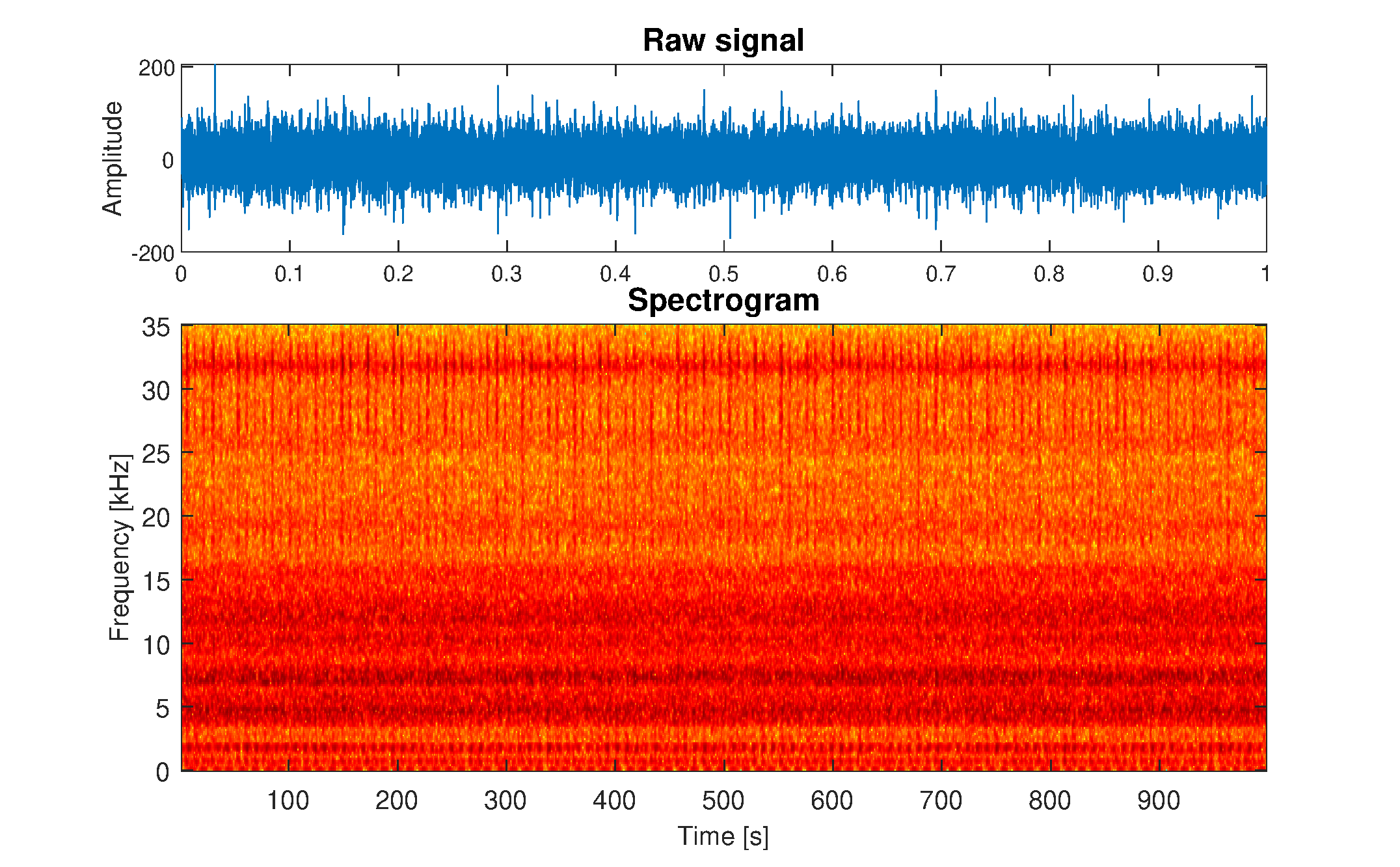}
\caption{Raw time series with spectrogram corresponding spectrogram.}\label{f:ac_CP_spec}
\end{figure}

The raw time series of the recorded signal from centrifugal pump and the corresponding spectrogram is presented in Figure \ref{f:ac_CP_spec}. In this case, the fault frequency is high i.e. 127 Hz. A high level of noise is observed in both the time series and the spectrogram. On the spectrogram, the pulses associated with the fault become visible for frequencies above 25 kHz. 

\begin{figure}[ht!]
\centering
\includegraphics[width=0.75\textwidth, angle=0]{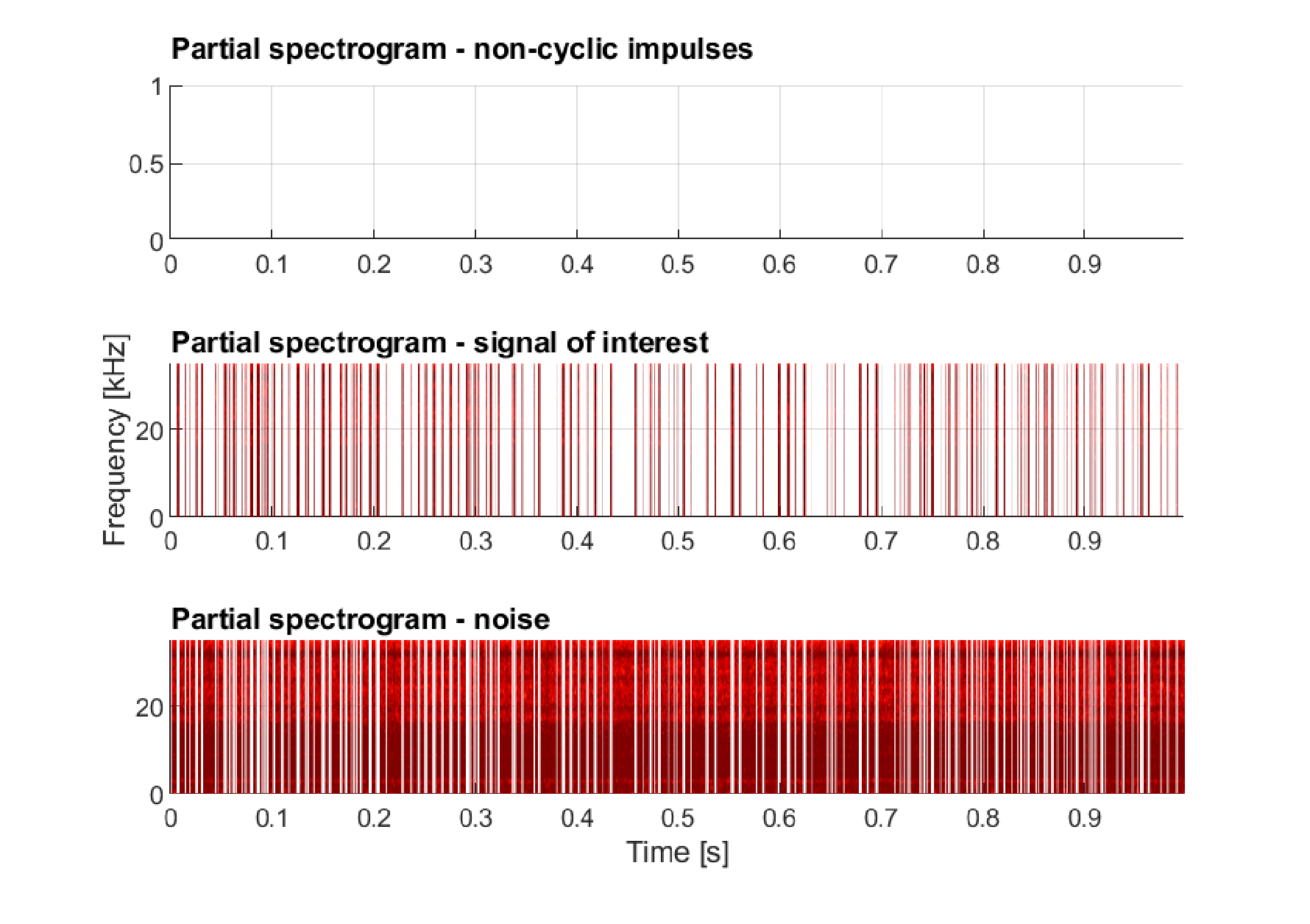}
\caption{Visual representation of DBSCAN clustering of the real acoustic data.}
\label{f:rl_cp_p}
\end{figure}

The partial spectrograms obtained in the first and second clustering are shown in Figure \ref{f:rl_cp_p}. The proposed algorithm performed as expected - no outlier observations were detected in the first step. The application of DBSCAN again allowed the detection of the signal of interest. After recovering the time series using MGLE, the resulting SES, along with the marked harmonics, is presented in the bottom panel of the Figure \ref{rl_ac_cp_comp_all_spectra}.

\begin{figure}[ht!]
\centering
\includegraphics[width=.7\textwidth, angle=0]{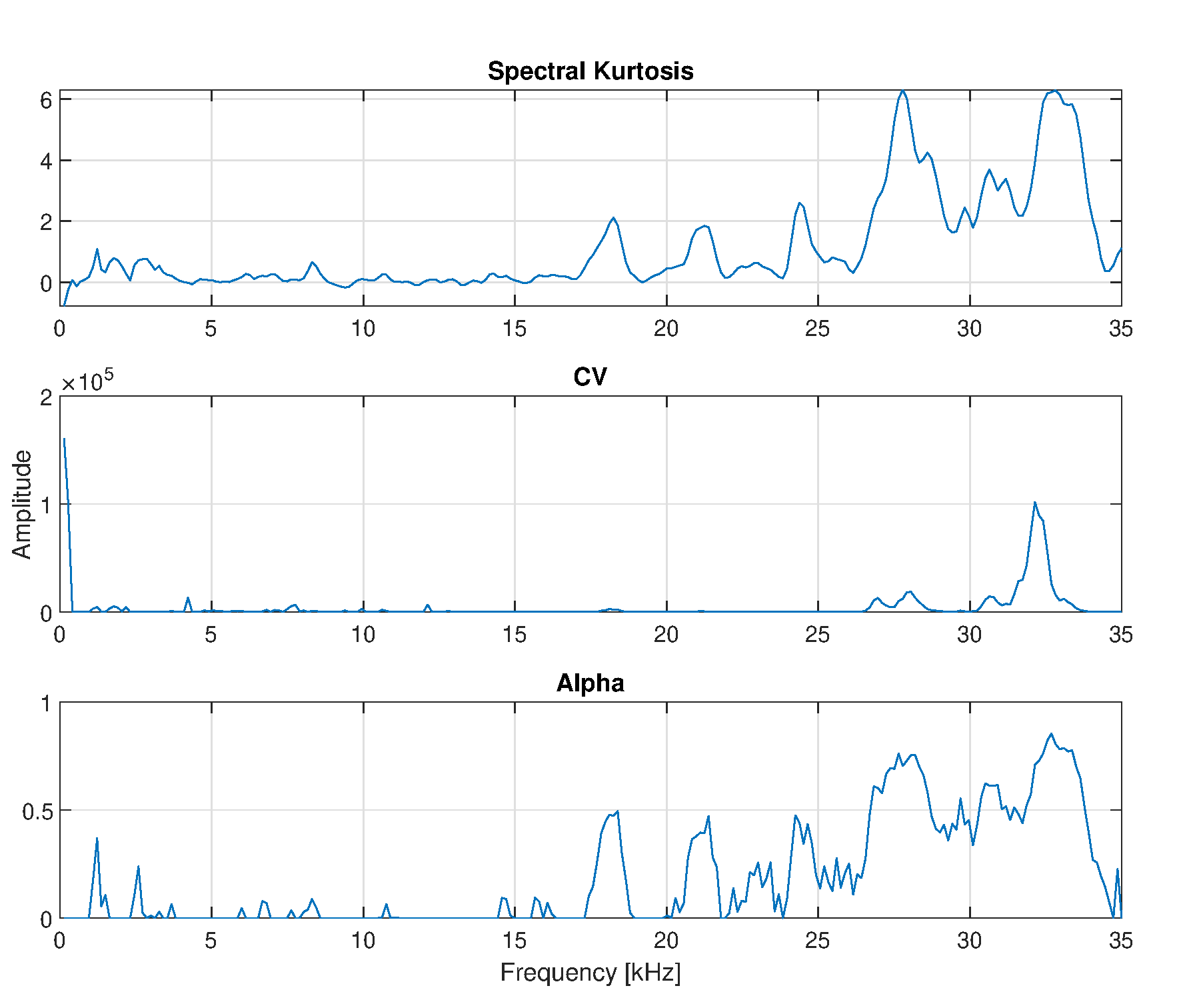}
\caption{Selectors used to comparison.}\label{f:rl_ac_cp_comp_selectors}
\end{figure}

Figure \ref{f:rl_ac_cp_comp_selectors} presents the selector results for the acoustic signal presented in Figure \ref{f:ac_CP_spec}. All of the compared selectors indicate higher values for frequencies above 25 kHz. However, the CV also indicates high values for very low frequencies, around zero. Due to this, the signal filtered with it exhibits the lowest ENVSI value (lower than the raw signal). The proposed approach yields results with similar ENVSI values to the other methods. However, the component corresponding to shaft rotation and its harmonics is less visible on the SES.

\begin{figure}[ht!]
\centering
\includegraphics[width=.75\textwidth, angle=0]{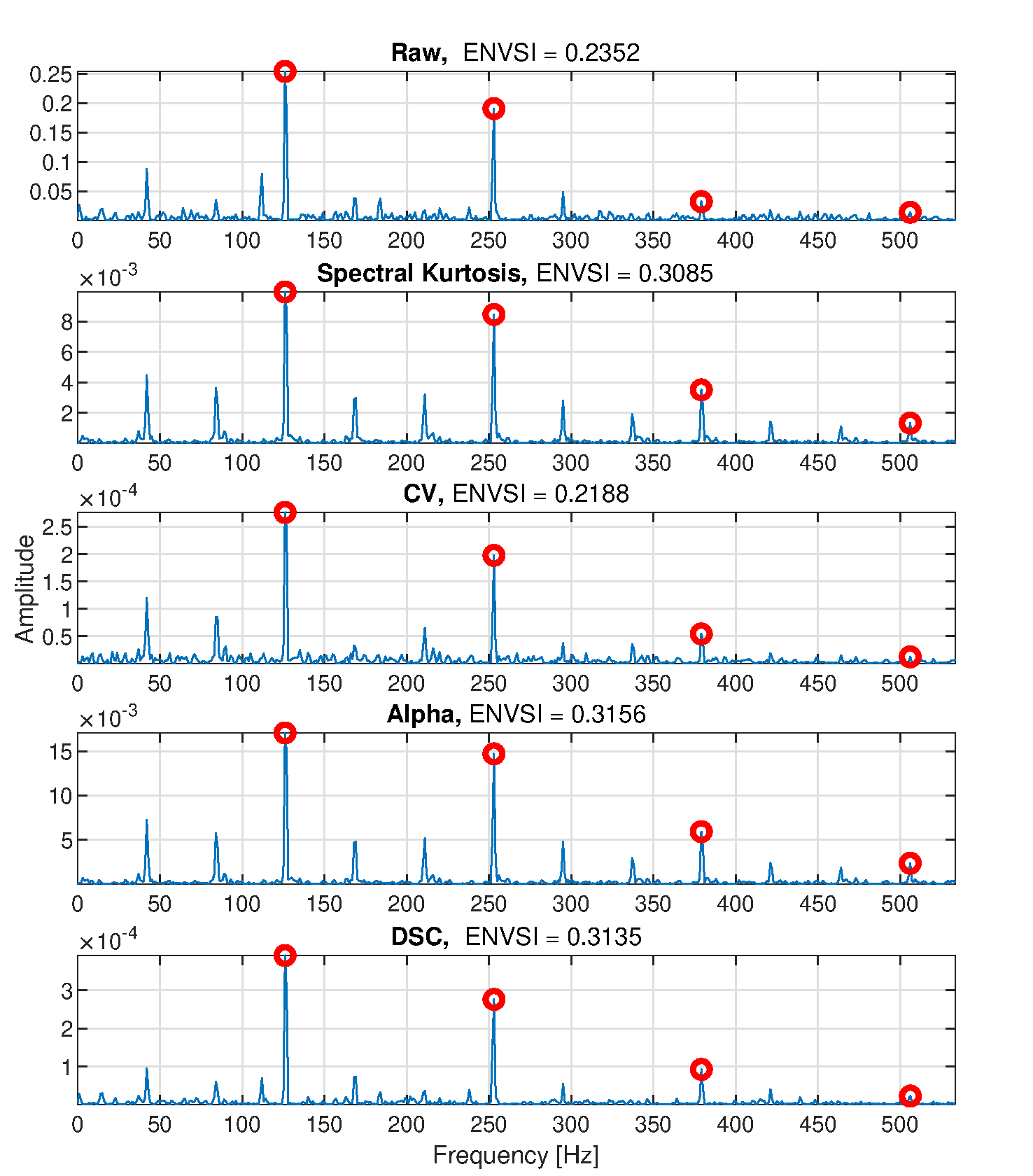}
\caption{The squared envelope spectra for the raw signal and all compared method.}\label{rl_ac_cp_comp_all_spectra}
\end{figure}

\subsubsection{Belt conveyor: faulty bearing in the idler}
\begin{figure}[ht!]
\centering
\includegraphics[width=.5\textwidth, angle=0]{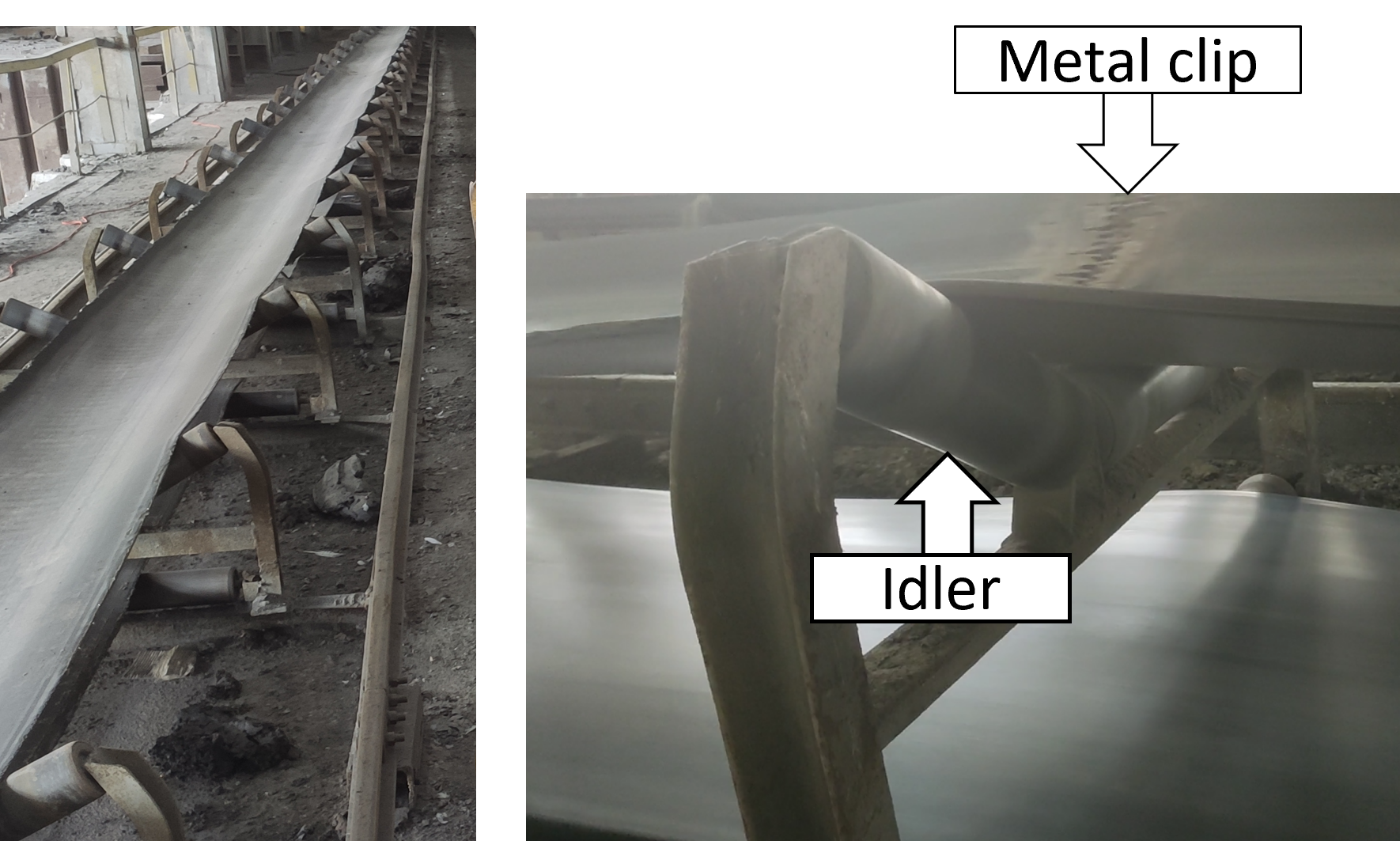}
\caption{The belt conveyor.}\label{f:idl_meas}
\end{figure}

 Idlers are parts of a belt conveyor that support the moving belt. Each idler has a shaft, two bearings, and a coating. Due to the large number of idlers involved, a fast, contactless method is required to assess the condition of the bearings. The experiment was carried out on a real industrial conveyor during regular operation (see Figure \ref{f:idl_meas}). Up to now, inspections have been carried out traditionally, where an expert examines the condition using their own senses, primarily sight and hearing. Throughout the experiments, the sound produced by each idler was recorded using a mobile phone. Specifically, the acoustic signals were extracted from videos taken with a typical smartphone. During the experiment, 10 s measurements were made for each idler with the sampling frequency $f_s = 48$ kHz. Several interesting instances were observed, including some idlers generating cyclic impulses in the signal, which correspond to the faulty bearing; however, it was discovered that significant disturbances occurred in some measurements.

\begin{figure}[ht!]
\centering
\includegraphics[width=.8\textwidth, angle=0]{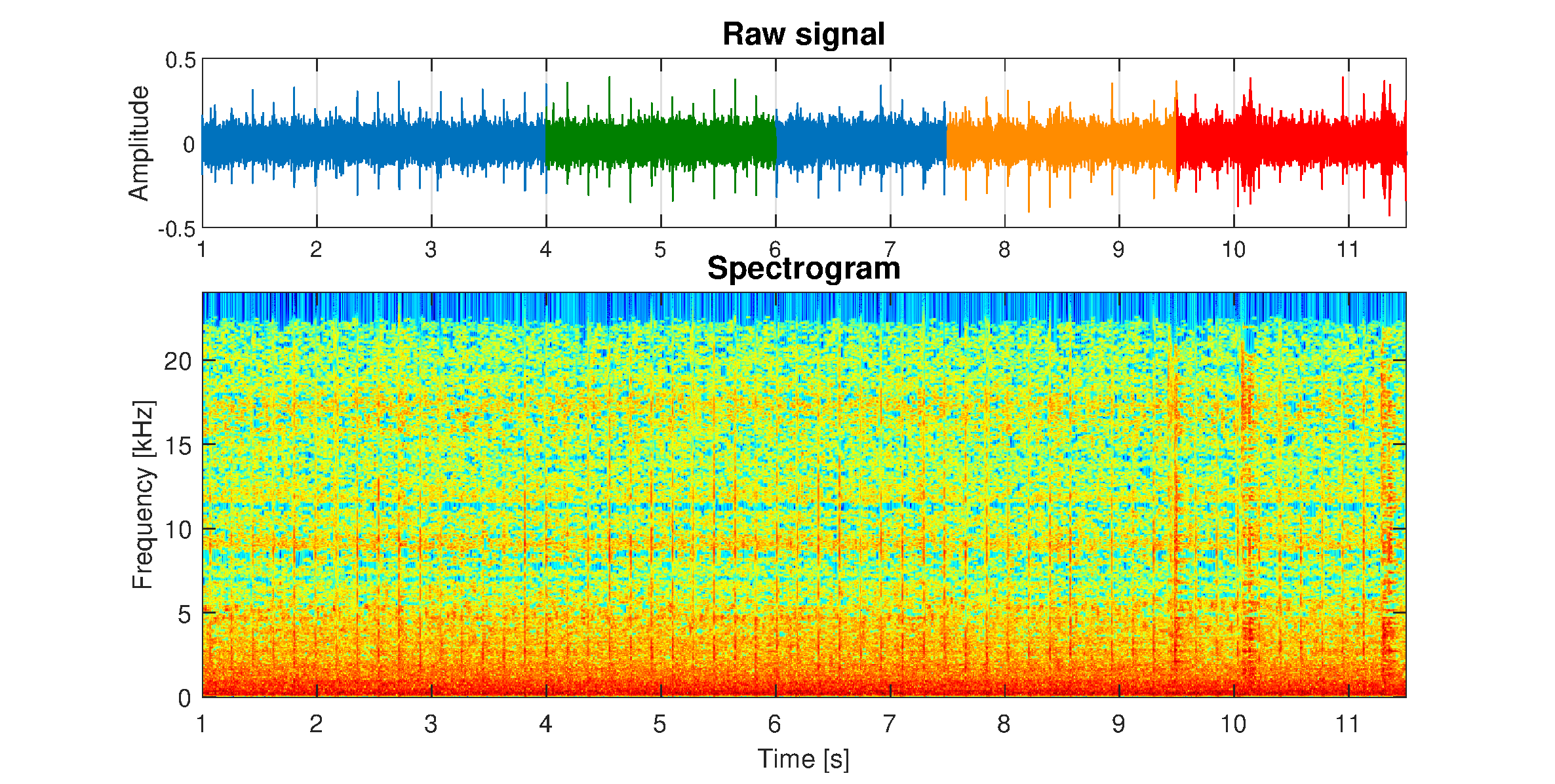}
\caption{Raw time series with spectrogram corresponding spectrogram.}\label{f:ac_spec}
\end{figure}

The raw time series of the recorded signal and the corresponding spectrogram is presented in Figure \ref{f:ac_spec}. Non-cyclic impulses, which make damage detection difficult, can be observed in both the time series and the spectrogram. These are three broadband impulses associated with the impact of a metal clip on the idler.

To evaluate the proposed method, the signal was divided into three two-second parts of varying difficulty. 

\begin{itemize}
    \item The first part, between 4 and 6 seconds, is labeled Case 2. It does not contain non-cyclic disturbances, and the damage is clearly visible above the noise level. In Figure \ref{f:ac_spec}, it is marked by green. It is a straightforward situation, and all the methods being compared should produce high-quality results. 
    \item The second part, between 7.5 and 9.5 seconds, is labeled Case 3. The random disturbances are still not appearing. In Figure \ref{f:ac_spec}, it is marked in orange. However, some challenges arise - two cyclic impulses are hidden in the noise, making them less visible on the spectrogram. 
    \item The third part, between 9.5 and 11.5 seconds, is labeled Case 4. This is the worst case. In Figure \ref{f:ac_spec}, this part of the signal is marked by red. The presence of two random broadband pulses in this fragment makes fault detection impossible for methods based only on frequency domain analysis. This situation highlights the advantage of the presented method.
\end{itemize}

\begin{figure}[ht!]
\centering
\begin{subfigure}[b]{0.32\textwidth}
\includegraphics[width=\textwidth, angle=0]{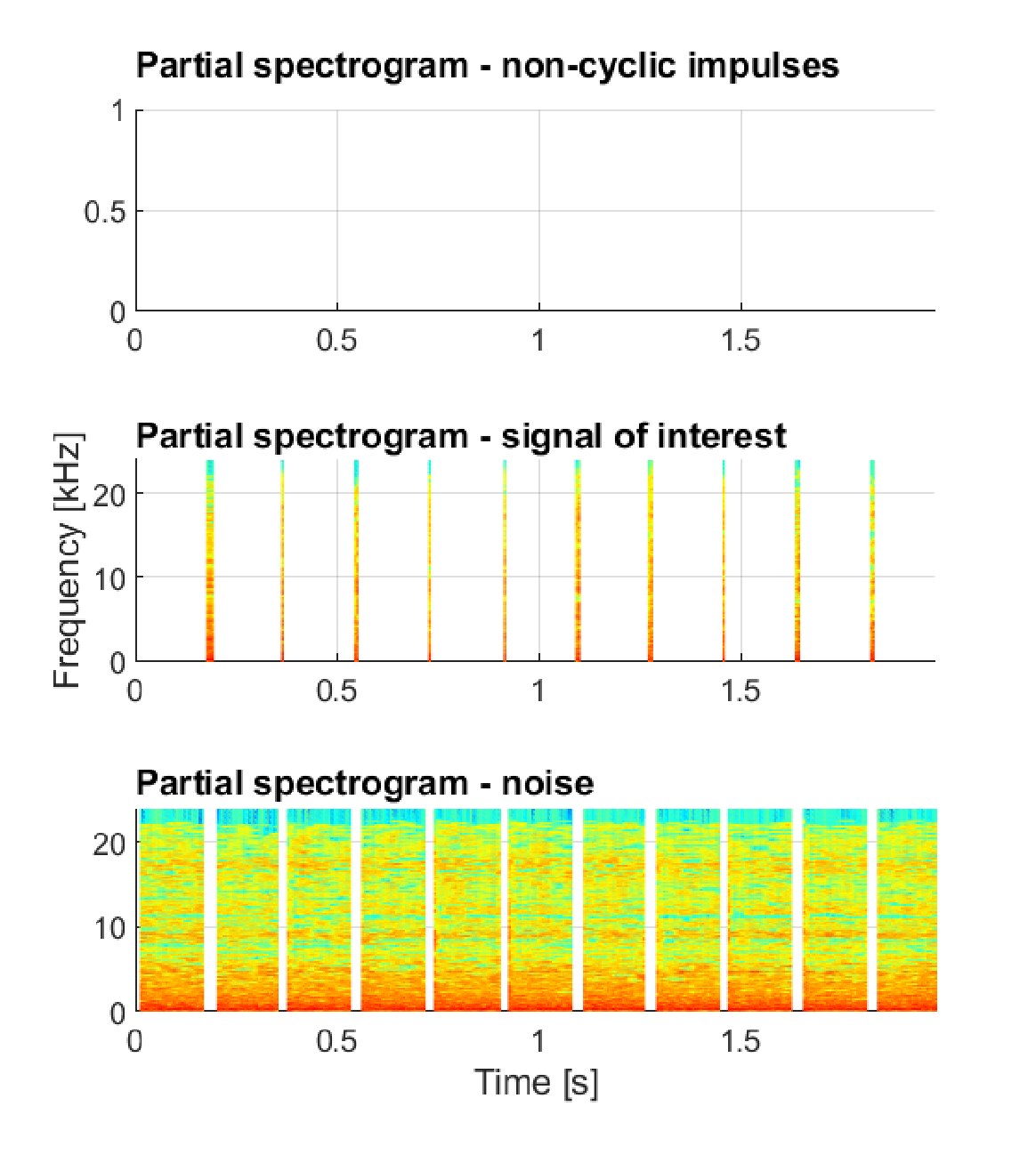}
\caption{Case 2}\label{f:rl_c2_p}
\end{subfigure}
\begin{subfigure}[b]{0.32\textwidth}
\includegraphics[width=\textwidth, angle=0]{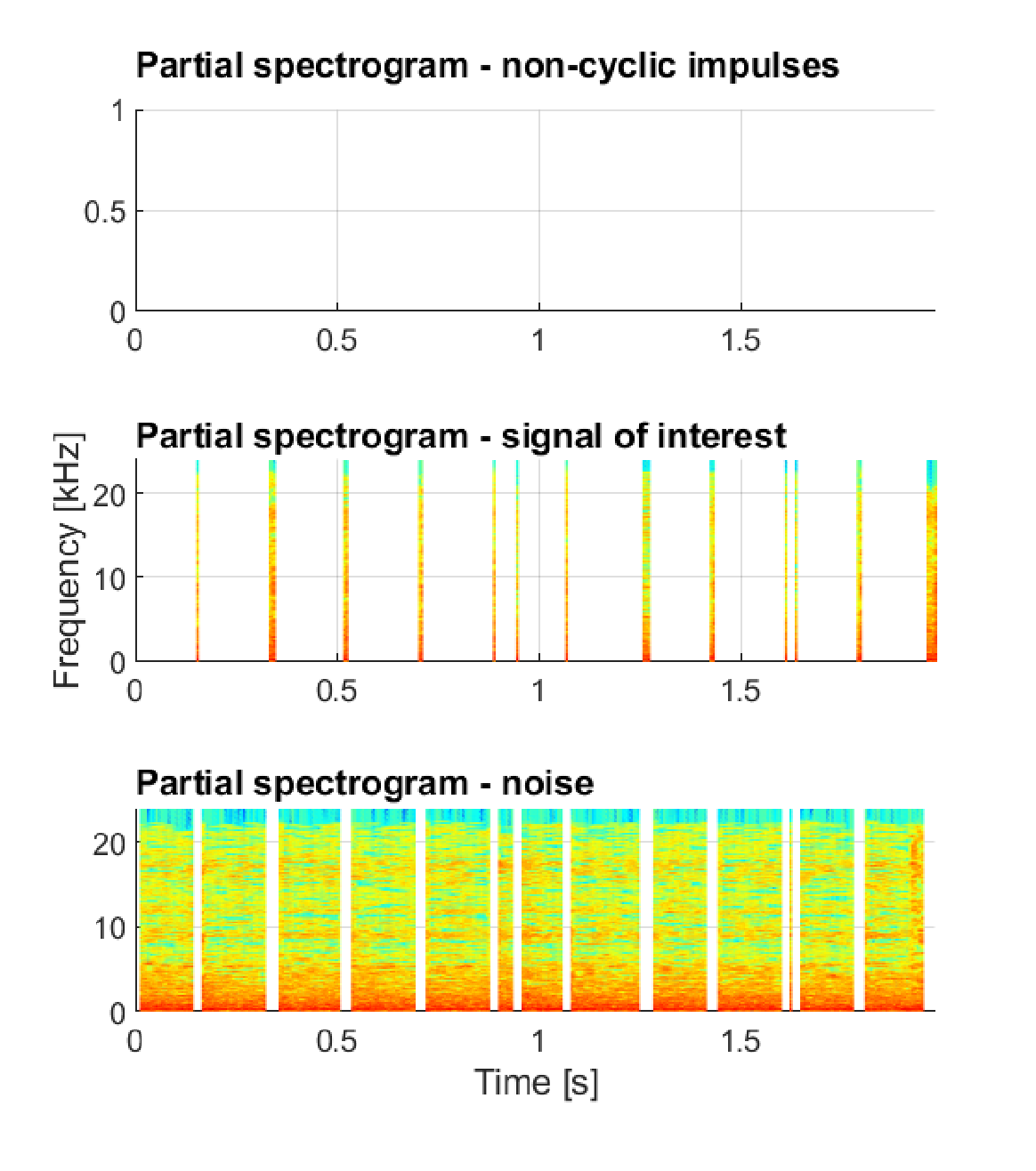}
\caption{Case 3}\label{f:rl_c3_p}
\end{subfigure}
\begin{subfigure}[b]{0.32\textwidth}
\includegraphics[width=\textwidth, angle=0]{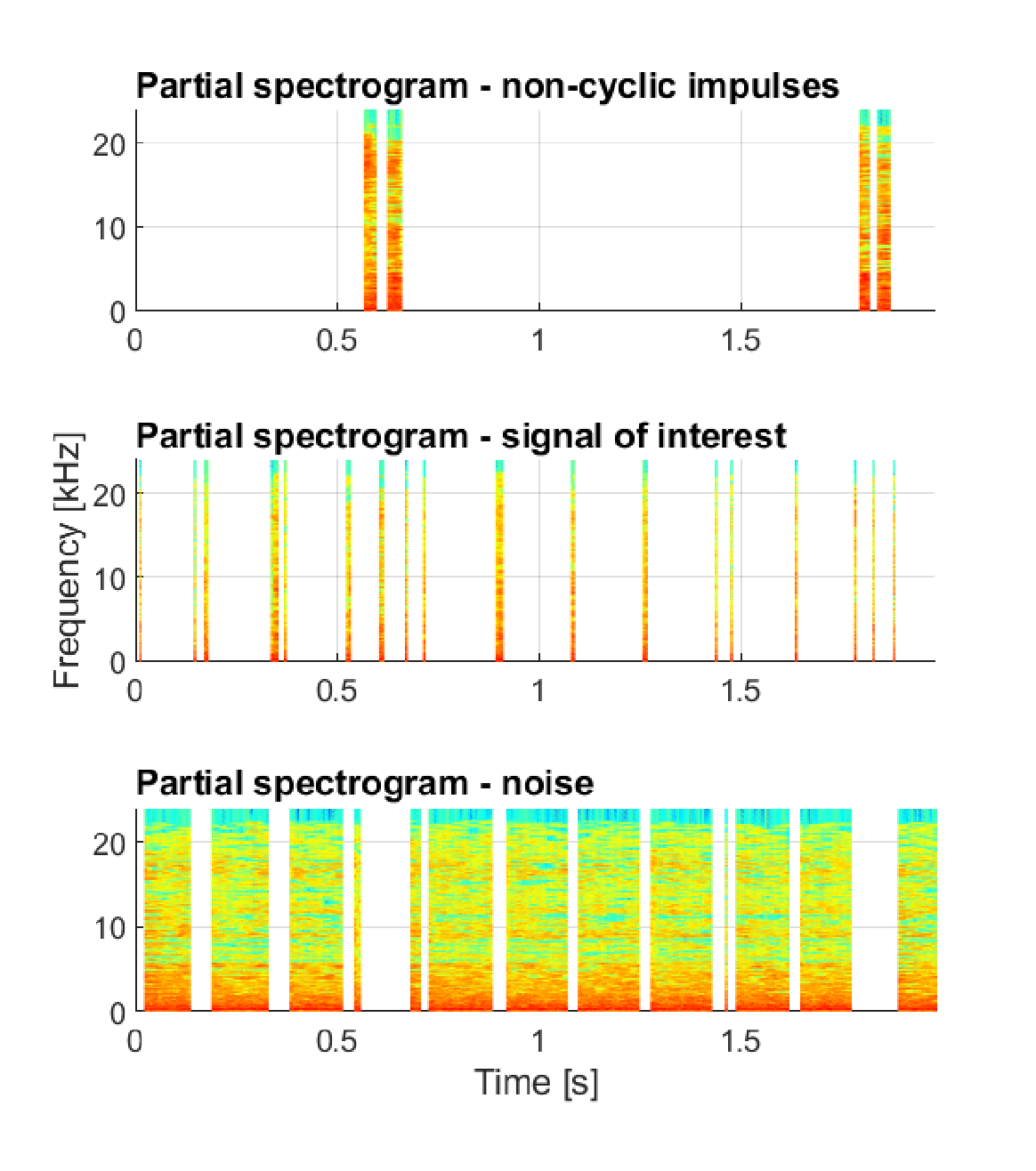}
\caption{Case 4}\label{f:rl_c4_p}
\end{subfigure}
\caption{Visual representation of DBSCAN clustering of the real acoustic data.}
\label{f:rl_ild_p}
\end{figure}

The partial spectrograms obtained by the first and second clustering stages are presented in Figure \ref{f:rl_ild_p}. The proposed algorithm effectively separates the classes required for further analysis. Non-cyclic disturbances were not detected in Cases 2 and 3. However, in Case 4, as expected, two impulses were separated during the initial clustering process (see the top panel of Figures \ref{f:rl_ild_p} (a-c)). As can be seen, class two, corresponding to the signal of interest, correctly identifies cyclic impulses in all three cases.

\begin{figure}[ht!]
\centering
\includegraphics[width=.8\textwidth, angle=0]{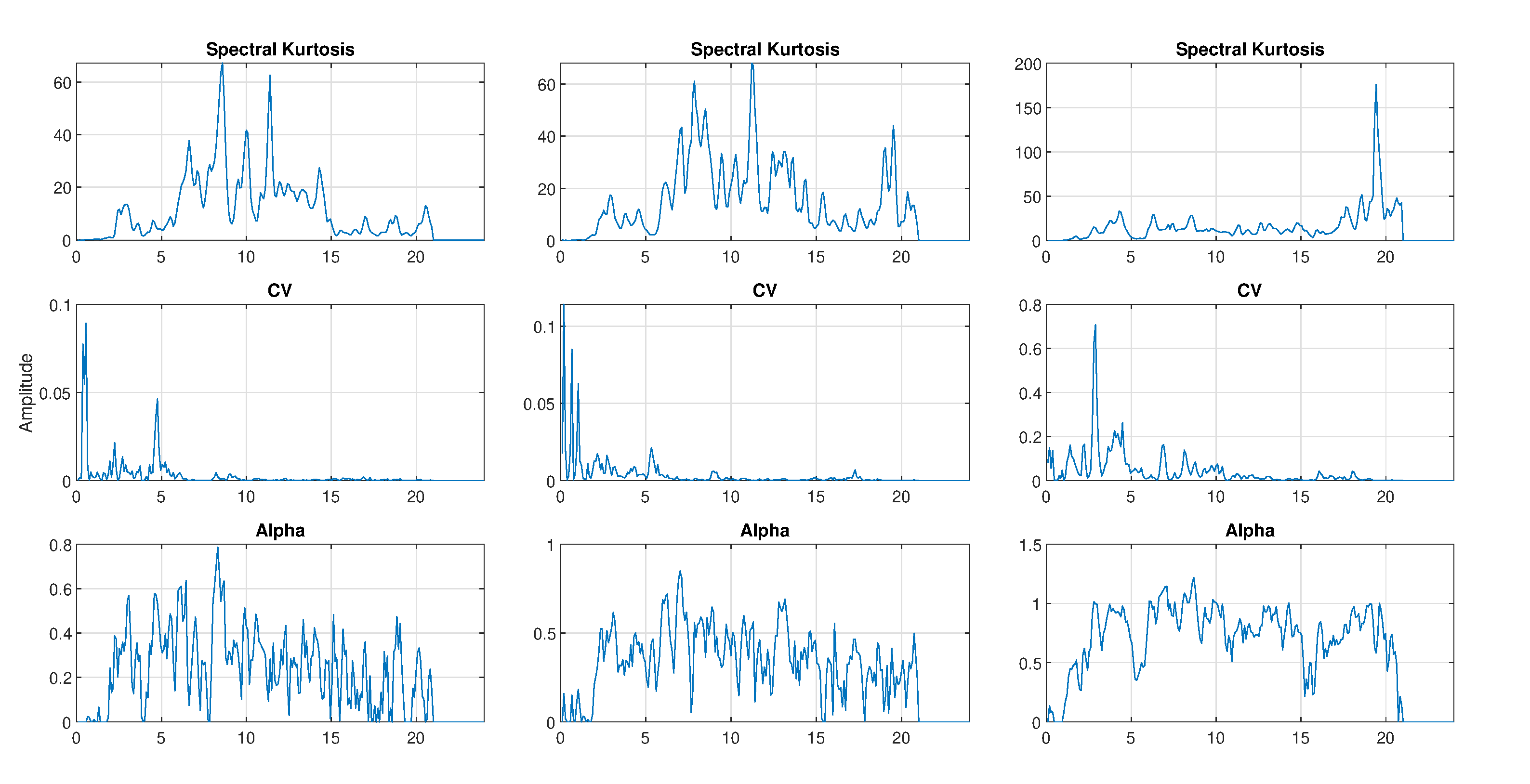}
\caption{Selectors used to comparison.}\label{f:rl_ac_idl_comp_selectors}
\end{figure} 

Figure \ref{f:rl_ac_idl_comp_selectors} demonstrates the selector results for the different fragments of signal analyzed. The first column shows the results for Case 2, the second column for Case 3, and the third column for Case 4. Three selectors were chosen for comparison, as in the case of the centrifugal pump: the classical spectral kurtosis and two additional ones dedicated to signals with non-Gaussian noise, which, unlike Case 1, is observed in the analyzed signal.

In Cases 2 and 3, SK and Alpha performed well in choosing the information band. However, in Case 4, the presence of non-cyclic disturbances in the same band as the cyclic impulses related to the fault presents a significant challenge. Even when alpha correctly identifies the informative frequency band, it is not possible to obtain information about the fault. The primary advantage of the method proposed by the authors is evident in this scenario. Time-domain analysis enables for accurate detection of damage when filtration methods fail.

\begin{figure}[ht!]
\centering
\includegraphics[width=\textwidth, angle=0]{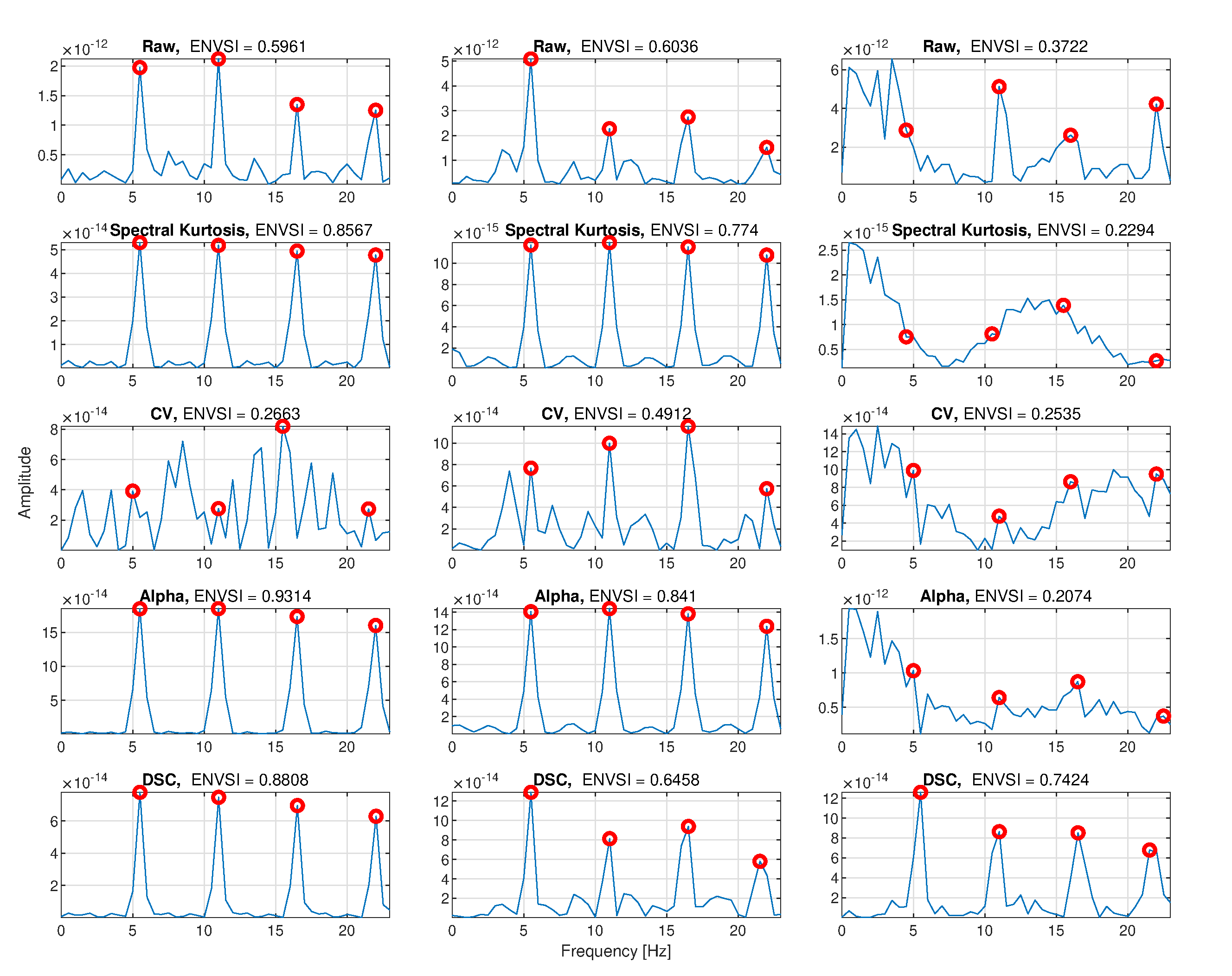}
\caption{The squared envelope spectra for the raw signal and all compared method.}\label{rl_ac_idl_comp_all_spectra}
\end{figure}

The squared envelope spectra of the recovered time series are shown in Figure \ref{rl_ac_idl_comp_all_spectra}. Red circles mark the harmonics corresponding to the fault frequency. The SES of the raw acoustic signal is presented in the first row. Next, the results obtained by the three selectors used for comparison, and finally, the results received by the proposed approach. With the exception of CV, in Cases 2 and 3, all methods correctly indicate the received damage frequency, that is, $5.47$ Hz. However, in case 4, the only method that indicates the correct result is the proposed DSC method. 

\begin{figure}[ht!]
\centering
\includegraphics[width=\textwidth, angle=0]{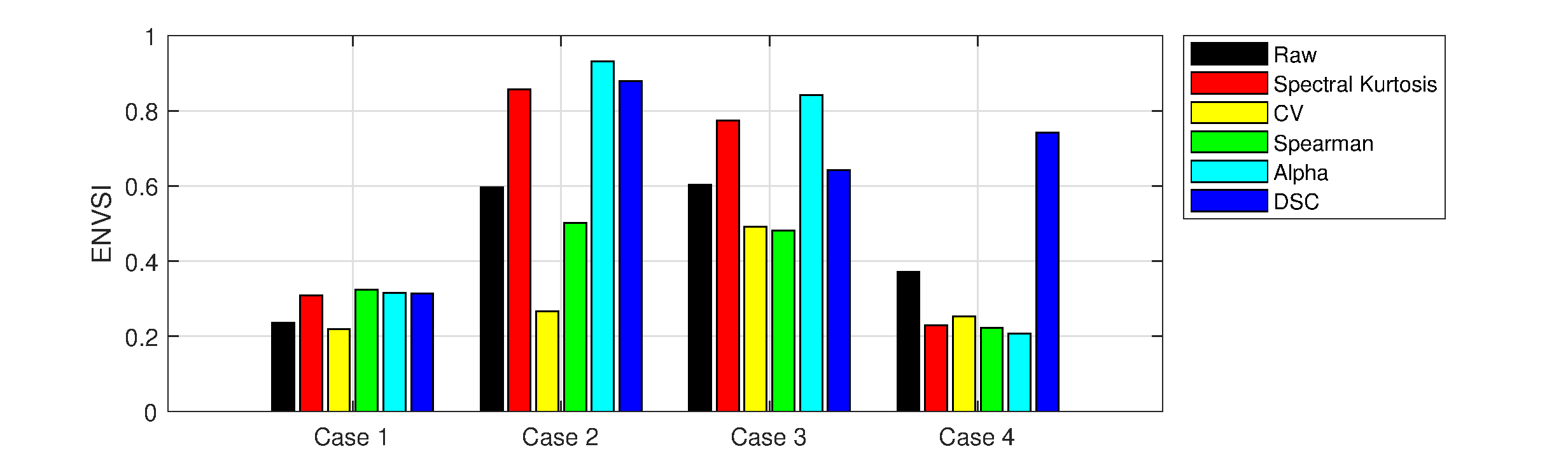}
\caption{The ENVSI value for the raw signal and all compared method}\label{rl_data_comp_all}
\end{figure}

For better visibility, the ENVSI values for all cases and methods analyzed are presented in bar plot (see Figure \ref{rl_data_comp_all}). If damage is visible in the signal, the ENVSI value obtained by the DSC is higher than that derived from the raw data, although not always exceeding the values obtained by the selectors. Note that it still allows one to detect the damage. The scenario changes when broadband non-cyclic impulses complicate the detection of damage (Case 4). Then, the proposed method yields satisfactory results when other methods fail. 


%


\section{Discussion}\label{sec4}

To evaluate the effectiveness of the presented algorithm, Monte Carlo simulations were conducted. 100 signals were simulated and analyzed for each of the 6 noise levels ($\sigma$) and 7 non-Gaussianity levels ($\gamma$). The parameters $\gamma$ and $\sigma$, as discussed in Section \ref{subsec3.1}, are associated with the behavior of the signal components. $\sigma$ represents the degree of concealment of the SOI component within the noise. A higher value of this parameter indicates a greater concealment of cyclic impulses within the background signal. The parameter $\gamma$ modifies the level of non-Gaussianity of the simulated signal. This variable directly affects the expected number of non-cyclic high-energy disturbances each second.

The effectiveness of the obtained results is presented from two perspectives. Firstly, an analysis of ENVSI is presented, which determines whether a family of components with fault frequency and harmonics have been correctly detected. 

\begin{figure}[ht!]
\centering
\includegraphics[width=0.6\textwidth]{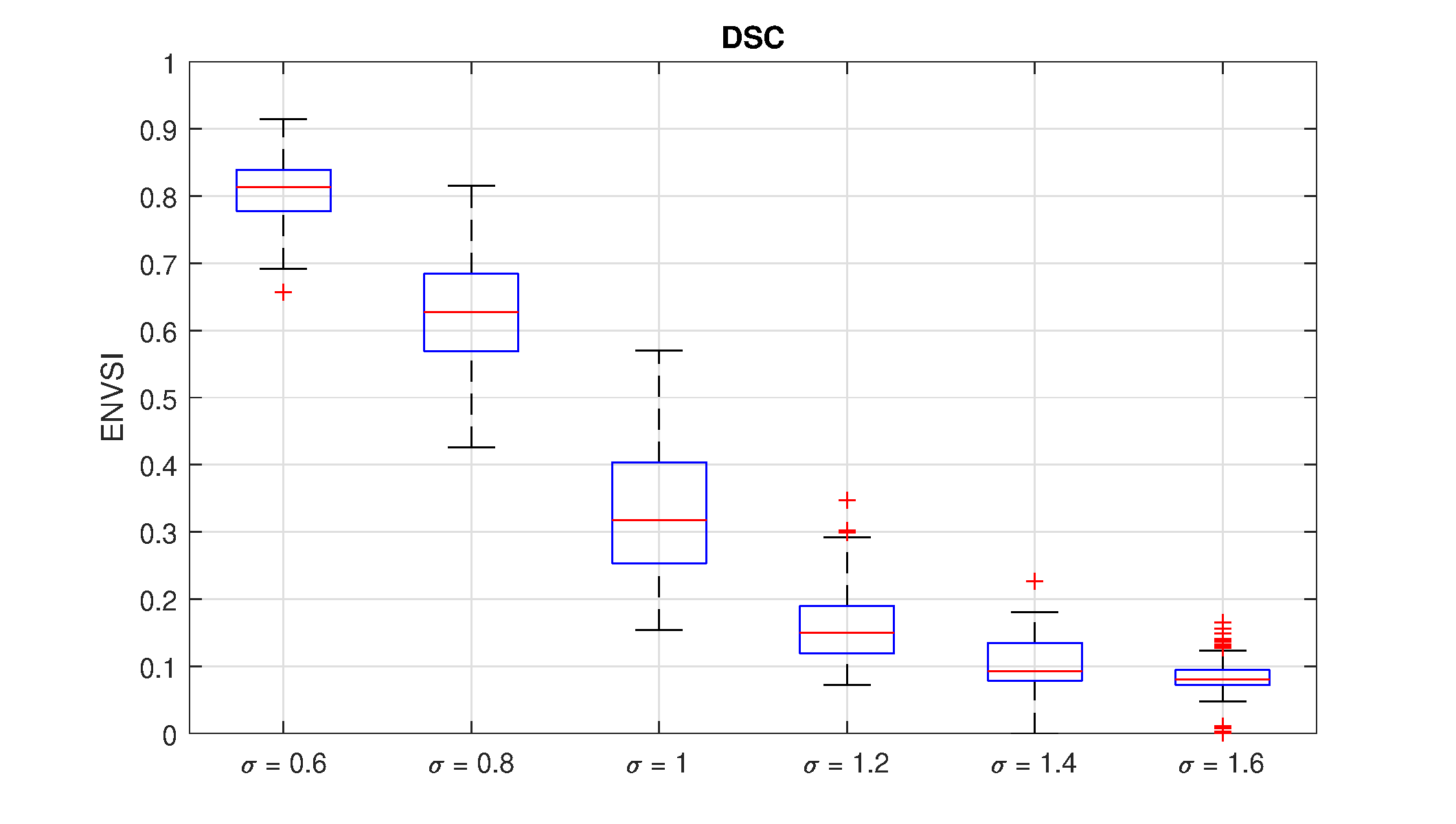}
\caption{Boxplot of the ENVSI indicator calculated based on Monte Carlo simulation for 100 iterations of the cyclic component (Class 2) for several levels of the noise (expressed as $\sigma$).}\label{envsi_MC}
\end{figure}

In Figure \ref{envsi_MC}, the behavior of ENVSI is observed depending on the simulation parameters. The red horizontal line denotes the median of 100 MC repetitions. If the median does not appear at the center of the box, it presents the sample skewness. The upper and lower edges of the box denote the upper and lower quartiles, respectively. The whiskers represent the most extreme data, which are not considered outliers. Outliers, marked by the red plus sign, are values that satisfy condition $1.5*IQR$, where IQR is the interquartile range. Due to the fact that no influence of the $\gamma$ parameter on the effectiveness of the method was observed, the results are presented only for $\gamma$ = 3. As expected, for low values of the $\sigma$ parameter, the median value is high (0.82). However, as $\sigma$ increases, indicating a higher level of concealment of the SOI within the noise, the effectiveness of the method decreases. In the case of $\sigma$ = 1.6, the results obtained are not satisfactory. The value of ENVSI is the same as that of the raw signal.

Secondly, the recovered signal corresponding to the SOI (Class 2) is considered. It is examined if the method allows the correct identification of the fault frequency.
As can be observed, when $\sigma \in [0.6, 1]$ the effectiveness of the proposed algorithm is 100\% (or almost 100\%). The hiding parameter in this range corresponds to the situation when the SOI component is noticed in the time series of the analyzed signal.

The number of non-cyclic impulses is directly determined by the parameter $\gamma$. This value corresponds to the expected number of impulses per second of the simulated signal. In the experiment presented, signals for which $\gamma$ is in the range of 1 to 7 were analyzed. In a situation where $\sigma \leq 1$ the compaction of the non-cyclic component does not significantly reduce the effectiveness of the proposed method.
If the estimated effectiveness of a method exceeds $90\%$, it is considered successful. It is noticeable that if $\sigma$ exceeds the value of 1, the efficiency of the presented method decreases. This means that when the SOI exceeds the amplitude of the noise component, the method loses its effectiveness. Then the result cannot be treated as a damage indicator. 

\begin{figure}
\centering
    \includegraphics[width=0.7\textwidth]{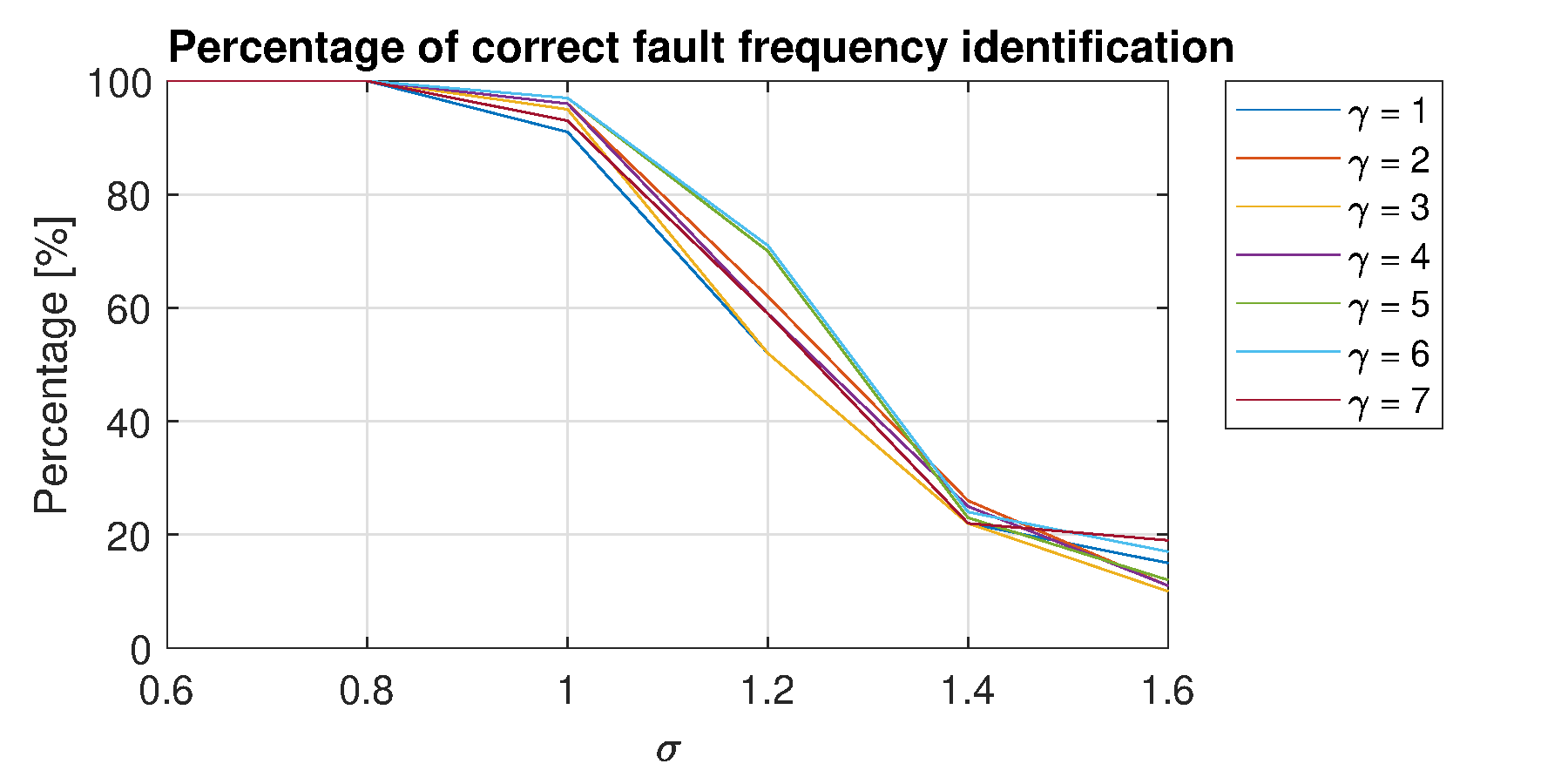}
    \caption{The percentage of correctly calculated SOI component frequencies.}
    \label{perc_MC}
\end{figure}

For signals with a similar noise amplitude and with an increase in the density of the non-cyclic component, the effectiveness of the proposed method in the analyzed range is similar.
To verify the accuracy of the algorithm, the input frequency of the SOI component was compared with the frequency obtained as a result of the analysis. The result with a precision of $\pm 0.5$ Hz in comparison to the input frequency was assumed to be the correct result. The results obtained are shown in Figure \ref{perc_MC}.


\section{Summary and conclusions}\label{sec5}

In this paper, the effectiveness of a method for separating components in a signal with non-Gaussian noise by analyzing spectrograms in the time domain and employing the DBSCAN algorithm for clustering is demonstrated. The results highlighte the method's effectiveness, particularly in noisy scenarios, where spatial clustering based on double density facilitated accurate identification of the signal of interest.

Monte Carlo simulations were performed to evaluate the effectiveness of the algorithm proposed in this study. Each simulation included 100 iterations for 6 noise levels ($\sigma$) and 7 non-Gaussianity levels ($\gamma$).
The effectiveness of the results is evaluated from two perspectives. Firstly, the ENVSI indicator was analyzed to detect the fault frequency accurately. Secondly, the recovered signal corresponding to the SOI (class 2) was examined to assess the method's capability of identifying the fault frequency. The findings revealed that for $\sigma \in [0.6, 1]$, the proposed algorithm achieved nearly perfect effectiveness. Furthermore, the density of non-cyclic impulses, as influenced by $\gamma$, showed only a marginal impact on the effectiveness of the method within the range analyzed.

The algorithm proved successful when applied to both simulated and real signals. Despite some challenges in detecting every impulse of the cyclic component due to spectrogram visibility issues, the general characteristics of the component were preserved. 
Additionally, the envelope spectrum proved useful in identifying characteristic frequencies, and the isolation of non-cyclic impulses was achieved.

Further work assumes the analysis and evaluation of the use of different methods for class separation. This will include different clustering algorithms, as well as other blind and non-blind methods.


\section*{Acknowledgements}

The work is supported by National Center of Science under Sheng2 project No. UMO- 2021/40/Q/ST8/00024 "NonGauMech - New methods of processing non-stationary signals (identification, segmentation, extraction, modeling) with non-Gaussian characteristics for the purpose of monitoring complex mechanical structures".





\bibliographystyle{elsarticle-num}\biboptions{sort&compress}
\bibliography{bibfile}

\begin{thebibliography}{10}
\expandafter\ifx\csname url\endcsname\relax
  \def\url#1{\texttt{#1}}\fi
\expandafter\ifx\csname urlprefix\endcsname\relax\def\urlprefix{URL }\fi
\expandafter\ifx\csname href\endcsname\relax
  \def\href#1#2{#2} \def\path#1{#1}\fi

\bibitem{zoubir2017contribution}
A.~Zoubir, N.~Yamina, A.~Noura, Contribution to the maintenance of t4 bh drilling machine (case of the mine of boukhadra, algeria), Mining Science 24 (2017) 73--83.

\bibitem{wodecki2020separation}
J.~Wodecki, A.~Michalak, R.~Zimroz, A.~Wy{\l}oma{\'n}ska, Separation of multiple local-damage-related components from vibration data using nonnegative matrix factorization and multichannel data fusion, Mechanical Systems and Signal Processing 145 (2020) 106954.

\bibitem{michalak2019application}
A.~Michalak, J.~Wodecki, A.~Wy{\l}oma{\'n}ska, R.~Zimroz, Application of cointegration to vibration signal for local damage detection in gearboxes, Applied Acoustics 144 (2019) 4--10.

\bibitem{yan2014wavelets}
R.~Yan, R.~X. Gao, X.~Chen, Wavelets for fault diagnosis of rotary machines: A review with applications, Signal processing 96 (2014) 1--15.

\bibitem{anwarsha2022review}
A.~Anwarsha, T.~Narendiranath~Babu, A review on the role of tunable q-factor wavelet transform in fault diagnosis of rolling element bearings, Journal of Vibration Engineering \& Technologies 10~(5) (2022) 1793--1808.

\bibitem{he2022bearing}
F.~He, Q.~Ye, A bearing fault diagnosis method based on wavelet packet transform and convolutional neural network optimized by simulated annealing algorithm, Sensors 22~(4) (2022) 1410.

\bibitem{liang2022intelligent}
P.~Liang, W.~Wang, X.~Yuan, S.~Liu, L.~Zhang, Y.~Cheng, Intelligent fault diagnosis of rolling bearing based on wavelet transform and improved resnet under noisy labels and environment, Engineering Applications of Artificial Intelligence 115 (2022) 105269.

\bibitem{antoni2007cyclic}
J.~Antoni, Cyclic spectral analysis of rolling-element bearing signals: facts and fictions, Journal of Sound and Vibration 304~(3) (2007) 497--529.

\bibitem{antoni2006spectral}
J.~Antoni, R.~B. Randall, The spectral kurtosis: application to the vibratory surveillance and diagnostics of rotating machines, Mechanical systems and signal processing 20~(2) (2006) 308--331.

\bibitem{hebda2020informative}
J.~Hebda-Sobkowicz, R.~Zimroz, M.~Pitera, A.~Wy{\l}oma{\'n}ska, Informative frequency band selection in the presence of non-gaussian noise--a novel approach based on the conditional variance statistic with application to bearing fault diagnosis, Mechanical Systems and Signal Processing 145 (2020) 106971.

\bibitem{obuchowski2014selection}
J.~Obuchowski, A.~Wy{\l}oma{\'n}ska, R.~Zimroz, Selection of informative frequency band in local damage detection in rotating machinery, Mechanical Systems and Signal Processing 48~(1-2) (2014) 138--152.

\bibitem{nowicki2021dependency}
J.~Nowicki, J.~Hebda-Sobkowicz, R.~Zimroz, A.~Wy{\l}oma{\'n}ska, Dependency measures for the diagnosis of local faults in application to the heavy-tailed vibration signal, Applied Acoustics 178 (2021) 107974.

\bibitem{antoni2007fast}
J.~Antoni, Fast computation of the kurtogram for the detection of transient faults, Mechanical Systems and Signal Processing 21~(1) (2007) 108--124.

\bibitem{antoni2016info}
J.~Antoni, The infogram: Entropic evidence of the signature of repetitive transients, Mechanical Systems and Signal Processing 74 (2016) 73--94.

\bibitem{zhao2016detection}
M.~Zhao, J.~Lin, Y.~Miao, X.~Xu, Detection and recovery of fault impulses via improved harmonic product spectrum and its application in defect size estimation of train bearings, Measurement 91 (2016) 421--439.

\bibitem{miao2017improvement_GINI}
Y.~Miao, M.~Zhao, J.~Lin, Improvement of kurtosis-guided-grams via gini index for bearing fault feature identification, Measurement Science and Technology 28~(12) (2017) 125001.

\bibitem{bozchalooi2007smoothness}
I.~S. Bozchalooi, M.~Liang, A smoothness index-guided approach to wavelet parameter selection in signal de-noising and fault detection, Journal of Sound and Vibration 308~(1-2) (2007) 246--267.

\bibitem{Moshrefzadeh2018294}
A.~Moshrefzadeh, A.~Fasana, {The Autogram: An effective approach for selecting the optimal demodulation band in rolling element bearings diagnosis}, Mechanical Systems and Signal Processing 105 (2018) 294 – 318.
\newblock \href {https://doi.org/10.1016/j.ymssp.2017.12.009} {\path{doi:10.1016/j.ymssp.2017.12.009}}.

\bibitem{Wang2022}
X.~Wang, J.~Zheng, Q.~Ni, H.~Pan, J.~Zhang, {Traversal index enhanced-gram (TIEgram): A novel optimal demodulation frequency band selection method for rolling bearing fault diagnosis under non-stationary operating conditions}, Mechanical Systems and Signal Processing 172 (2022).
\newblock \href {https://doi.org/10.1016/j.ymssp.2022.109017} {\path{doi:10.1016/j.ymssp.2022.109017}}.

\bibitem{schmidt2020methodology}
S.~Schmidt, A.~Mauricio, P.~S. Heyns, K.~C. Gryllias, A methodology for identifying information rich frequency bands for diagnostics of mechanical components-of-interest under time-varying operating conditions, Mechanical Systems and Signal Processing 142 (2020) 106739.

\bibitem{Liu2024}
W.~Liu, S.~Yang, Y.~Liu, X.~Gu, {DTMSgram: a novel optimal demodulation frequency band selection method for wheelset bearings fault diagnosis under wheel-rail excitation}, Measurement Science and Technology 35~(4) (2024).
\newblock \href {https://doi.org/10.1088/1361-6501/ad0d74} {\path{doi:10.1088/1361-6501/ad0d74}}.

\bibitem{pancaldi2023impact}
F.~Pancaldi, L.~Dibiase, M.~Cocconcelli, Impact of noise model on the performance of algorithms for fault diagnosis in rolling bearings, Mechanical Systems and Signal Processing 188 (2023) 109975.

\bibitem{saufi2017review}
M.~Saufi, Z.~A. Ahmad, M.~H. Lim, M.~Leong, A review on signal processing techniques for bearing diagnostics, Int J Mech Eng Technol 8~(6) (2017) 327--337.

\bibitem{michalak2020model}
A.~Michalak, J.~Wodecki, M.~Drozda, A.~Wy{\l}oma{\'n}ska, R.~Zimroz, Model of the vibration signal of the vibrating sieving screen suspension for condition monitoring purposes, Sensors 21~(1) (2020) 213.

\bibitem{smith2016optimised}
W.~A. Smith, Z.~Fan, Z.~Peng, H.~Li, R.~B. Randall, Optimised spectral kurtosis for bearing diagnostics under electromagnetic interference, Mechanical Systems and Signal Processing 75 (2016) 371--394.

\bibitem{kruczek2020detect}
P.~Kruczek, R.~Zimroz, A.~Wy{\l}oma{\'n}ska, How to detect the cyclostationarity in heavy-tailed distributed signals, Signal Processing 172 (2020) 107514.

\bibitem{wodecki2021influence}
J.~Wodecki, A.~Michalak, A.~Wy{\l}oma{\'n}ska, R.~Zimroz, Influence of non-gaussian noise on the effectiveness of cyclostationary analysis--simulations and real data analysis, Measurement 171 (2021) 108814.

\bibitem{wylomanska2016impulsive}
A.~Wy{\l}oma{\'n}ska, R.~Zimroz, J.~Janczura, J.~Obuchowski, Impulsive noise cancellation method for copper ore crusher vibration signals enhancement, IEEE Transactions on Industrial Electronics 63~(9) (2016) 5612--5621.

\bibitem{borghesani2017cs2}
P.~Borghesani, J.~Antoni, Cs2 analysis in presence of non-gaussian background noise--effect on traditional estimators and resilience of log-envelope indicators, Mechanical Systems and Signal Processing 90 (2017) 378--398.

\bibitem{barszcz2011novel}
T.~Barszcz, A.~Jab{\l}o{\'n}ski, A novel method for the optimal band selection for vibration signal demodulation and comparison with the kurtogram, Mechanical systems and signal processing 25~(1) (2011) 431--451.

\bibitem{WODECKI2021108400}
J.~Wodecki, A.~Michalak, R.~Zimroz, Local damage detection based on vibration data analysis in the presence of gaussian and heavy-tailed impulsive noise, Measurement 169 (2021) 108400.

\bibitem{zheng2024progressive}
X.~Zheng, Z.~He, J.~Nie, P.~Li, Z.~Dong, M.~Gao, A progressive multi-source domain adaptation method for bearing fault diagnosis, Applied Acoustics 216 (2024) 109797.

\bibitem{ye2023intelligent}
M.~Ye, X.~Yan, N.~Chen, M.~Jia, Intelligent fault diagnosis of rolling bearing using variational mode extraction and improved one-dimensional convolutional neural network, Applied Acoustics 202 (2023) 109143.

\bibitem{xu2022fault}
Q.~Xu, B.~Zhu, H.~Huo, Z.~Meng, J.~Li, F.~Fan, L.~Cao, Fault diagnosis of rolling bearing based on online transfer convolutional neural network, Applied Acoustics 192 (2022) 108703.

\bibitem{wissbrock2024more}
P.~Wi{\ss}brock, Z.~Ren, D.~Pelkmann, More than spectrograms: Deep representation learning for machinery fault detection, Applied Acoustics 225 (2024) 110178.

\bibitem{ester1996density}
M.~Ester, H.-P. Kriegel, J.~Sander, X.~Xu, et~al., A density-based algorithm for discovering clusters in large spatial databases with noise, in: Conference on Knowledge Discovery and Data Mining, Vol.~96, 1996, pp. 226--231.

\bibitem{alam1979estimation}
K.~Alam, Estimation of multinomial probabilities, The Annals of Statistics 7~(2) (1979) 282--283.

\bibitem{allen1977short}
J.~Allen, Short term spectral analysis, synthesis, and modification by discrete {F}ourier transform, IEEE Transactions on Acoustics, Speech, and Signal Processing 25~(3) (1977) 235--238.

\bibitem{griffin1984signal}
D.~Griffin, J.~Lim, Signal estimation from modified short-time fourier transform, IEEE Transactions on Acoustics, Speech, and Signal Processing 32~(2) (1984) 236--243.

\bibitem{Taqqu}
G.~Samoradnitsky, Stable Non-Gaussian Random Processes: Stochastic Models with Infinite Variance, Chapman \& Hall, 1994.

\bibitem{zak2015application}
G.~{\.Z}ak, A.~Wy{\l}oma{\'n}ska, R.~Zimroz, Application of alpha-stable distribution approach for local damage detection in rotating machines, Journal of Vibroengineering 17~(6) (2015) 2987--3002.

\bibitem{HebdziaSobkowicz2020b}
J.~Hebda-Sobkowicz, R.~Zimroz, M.~Pitera, A.~Wylomanska, Informative frequency band selection in the presence of non-gaussian noise - a novel approach based on the conditional variance statistic with application to bearing fault diagnosis, Mechanical Systems and Signal Processing 145 (2020) 106971.

\bibitem{jelito2021new}
D.~Jelito, M.~Pitera, New fat-tail normality test based on conditional second moments with applications to finance, Statistical Papers 62~(5) (2021) 2083--2108.

\end{thebibliography}






\end{document}